\documentclass[10pt,journal,letterpaper]{IEEEtran}

\usepackage{ifpdf}

\ifpdf
\usepackage[pdftex]{graphicx}
\usepackage{mathrsfs}
\usepackage{amsmath}
\usepackage{amssymb}
\usepackage{amsfonts,amssymb}
\usepackage[square, comma, sort&compress, numbers]{natbib}
\usepackage{algorithm}
\usepackage{algorithmicx}
\usepackage[noend]{algpseudocode}

\usepackage{enumerate}
\usepackage{multirow}
\usepackage{array}
\usepackage{tabularx}
\usepackage{makecell}
\usepackage{amsthm}
\usepackage{color}
\usepackage[numbers]{natbib}
\usepackage[square,numbers,sort&compress]{natbib}

\usepackage{lipsum} 
\newcommand{\tabincell}[2]{\begin{tabular}{@{}#1@{}}#2\end{tabular}}

\usepackage{array}
\usepackage{subcaption}

\newcommand{\PreserveBackslash}[1]{\let\temp=\\#1\let\\=\temp}
\newcolumntype{C}[1]{>{\PreserveBackslash\centering}p{#1}}
\newcolumntype{R}[1]{>{\PreserveBackslash\raggedleft}p{#1}}
\newcolumntype{L}[1]{>{\PreserveBackslash\raggedright}p{#1}}

\usepackage{marvosym}
\usepackage{CJK}
\usepackage{url}
\usepackage{multirow}
\usepackage{diagbox}
\usepackage{booktabs}

\else
\fi

\ifCLASSINFOpdf
\else
\fi

\begin{document}

\title{Correlating Account on Ethereum Mixing Service via Domain-Invariant feature learning
}
	
\author{ 
	Zheng Che, Taoyu Li, 
	Meng Shen,~\IEEEmembership{Member,~IEEE,} Hanbiao Du, Liehuang Zhu,~\IEEEmembership{Senior Member,~IEEE}
	
	\IEEEcompsocitemizethanks{
		\IEEEcompsocthanksitem Z. Che is with the School of Computer Science, Beijing Institute of Technology, Beijing 100081, China (e-mail: \ chezheng@bit.edu.cn).
        \IEEEcompsocthanksitem T. Li is with the School of Computer Science and Technology, Taiyuan University of Technology, JinZhong 030600, China (e-mail: \ 921034826@qq.com).
		\IEEEcompsocthanksitem M. Shen, H. Du and L. Zhu are with the School of Cyberspace Science and Technology, Beijing Institute of Technology, Beijing 100081, China (e-mail: \{shenmeng, duhanbiao, liehuangz\}@bit.edu.cn).
		
	}
}
	
\maketitle

\begin{abstract}
The untraceability of transactions facilitated by Ethereum mixing services like Tornado Cash poses significant challenges to blockchain security and financial regulation. Existing methods for correlating mixing accounts suffer from limited labeled data and vulnerability to noisy annotations, which restrict their practical applicability. In this paper, we propose StealthLink, a novel framework that addresses these limitations through cross-task domain-invariant feature learning. Our key innovation lies in transferring knowledge from the well-studied domain of blockchain anomaly detection to the data-scarce task of mixing transaction tracing. Specifically, we design a MixFusion module that constructs and encodes mixing subgraphs to capture local transactional patterns, while introducing a knowledge transfer mechanism that aligns discriminative features across domains through adversarial discrepancy minimization. This dual approach enables robust feature learning under label scarcity and distribution shifts. Extensive experiments on real-world mixing transaction datasets demonstrate that StealthLink achieves state-of-the-art performance, with 96.98\% F1-score in 10-shot learning scenarios. Notably, our framework shows superior generalization capability in imbalanced data conditions than conventional supervised methods. This work establishes the first systematic approach for cross-domain knowledge transfer in blockchain forensics, providing a practical solution for combating privacy-enhanced financial crimes in decentralized ecosystems.

\end{abstract}
               

%
\IEEEpeerreviewmaketitle

\section{Introduction}\IEEEPARstart{I}{n} recent years, Web3.0 has garnered considerable attention as a transformative paradigm for the internet \cite{web3}. With its decentralized and user-centric characteristics, Web3.0 enables users to independently manage their identity information, create digital works, and engage in digital asset transactions, thereby significantly facilitating the circulation of data value. Ethereum\cite{Ethereum}, as a vital blockchain platform underpinning the value circulation within the Web3.0 ecosystem, has also garnered significant attention from various stakeholders. According to statistical data from CoinMarketCap\footnote{\url{https://coinmarketcap.com/}}, the market capitalization of Ethereum's native token ETH exceeded \$220 billion as of April 2025.

As the most prominent mixing service provider in Ethereum, Tornado Cash (TC) \cite{TornadoCash:online} are designed to enhance user privacy by obfuscating the traceability of transactions, thus making it difficult to link specific coins with their previous owners. However, this technique has raised substantial concerns regarding illicit activities, including money laundering, as well as the potential for cryptocurrencies to be misused. In February 2024, the co-founder of the video game Axy Infinity was stolen with hackers stealing 3,248 ETH and sending them to Tornado Cash to evade traking \cite{hacker1:online}. In addition, Tornado Cash is accused of allegedly facilitating money laundering transactions amounting to almost \$1 billion on behalf of the criminal organization known as the Lazarus Group \cite{hacker2:online}. The untraceability afforded by Tornado Cash presents a significant menace to the blockchain ecosystem and financial stability. Consequently, there is an urgent imperative to dismantle the anonymity of Tornado Cash by establishing associations among the addresses involved in mixing transactions.

\begin{figure}[htbp]
	\small
	\centering
	\includegraphics[width=\linewidth]{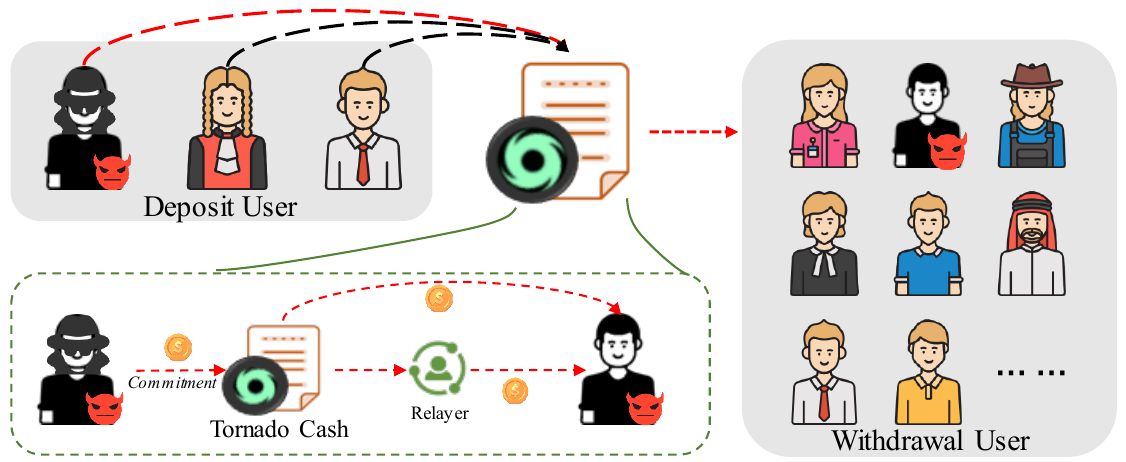}
	\caption{Tornado Cash is abused for illegal behavior.}
	\label{fig TC}
\end{figure}

Figure \ref{fig TC} illustrates the factors contributing to the misuse of TC for criminal activities. Criminals exploit TC by submitting self-generated zk-snark promises, allowing them to transfer illicit funds into the platform's pool. Subsequently, these funds are withdrawn into new accounts. Due to the substantial size of the TC withdrawal user base, the transfer of illegal funds to new accounts occurs discreetly, making it challenging for regulators to detect such illicit activities.

Recent studies have put forward various methods to identify mixing service addresses using traditional approaches, including heuristic rules \cite{DAPPS21,WWW2023} and empirical analysis \cite{Usenix2023,WWW2021}. However, these methods primarily rely on transaction relationships between user addresses to infer fund flow, without fully exploiting the intricate associations within address neighborhoods. Additionally, some methods based on machine learning, including deep learning \cite{zhengzibin2022,XiongGang2023} or graph neural networks \cite{Mixbroker,IOTJ23}  to detect mixing addresses. Nevertheless, these analytical techniques often necessitate extensive real-world datasets, and in the case of Tornado Cash, labeled mixing addresses are exceptionally scarce.

To address the challenges of transaction traceability caused by the abuse of mixing technologies, researchers have proposed various technical approaches for tracing mixed transactions, including identifying addresses controlled by mixing services \cite{Usenix2023,WWW2021} and correlating input-output addresses in mixed transactions \cite{DAPPS21,WWW2023}. However, these methods heavily rely on the construction of ground-truth datasets for mixed transactions, where the quality of such datasets directly determines the accuracy of traceability models. Furthermore, the limited scale of ground-truth datasets restricts supervised learning-based traceability methods \cite{zhengzibin2022,XiongGang2023}, as insufficient labeled data hinders the establishment of comprehensive decision boundaries, leading to overfitting tendencies in complex mixing scenarios and degraded generalization performance in real-world large-scale transaction environments.

This paper proposes StealthLink, a mixing transaction traceability framework that achieves high-precision tracing under limited labeled samples while enhancing model robustness by learning cross-task invariant features between abnormal blockchain transactions and mixed transactions. Specifically, to improve association discrimination in data-scarce scenarios, we design a MixFusion module (Mixing Subgraph Fusion Encoding). This module captures local transactional behaviors and structural patterns of mixing addresses by constructing mixing subgraphs and fuses their embedded representations to generate joint features for association discrimination, thereby transforming complex transaction correlation tasks into graph classification problems. To ensure robustness against noisy data, we introduce a knowledge transfer module that enables the model to learn consistent discriminative features across the domains of abnormal transactions and mixed transactions. This alignment of cross-domain representations enhances adaptability to varying data distributions and maintains stable performance in the presence of label noise.

The primary contributions of this chapter are summarized as follows:

\begin{itemize}
\item First systematic analysis of the feasibility of transferring knowledge from blockchain anomaly detection to mixing transaction traceability, revealing the potential advantages of cross-domain invariant features in addressing mixing-related anonymity challenges.
\item A novel framework, StealthLink, which leverages cross-task invariant feature learning to train highly robust and accurate traceability models using minimal labeled data.
\item Extensive experiments on mixing transaction datasets, including few-shot learning, robustness testing, and imbalanced data scenarios. Results demonstrate that StealthLink achieves state-of-the-art performance across all evaluated conditions.
\end{itemize}

The rest of this paper is organized as follows. 
We first introduce the background of Tornado Cash and summarize the related work in Section \ref{sec:Related Work}. Then, we describe the design goals in Section \ref{sec:Formal definition}. We introduce the transaction representation in Section IV and present StealthLink in Section \ref{sec:StealthLink}. Next, We evaluate the performance of StealthLink and compare it comprehensively with the state-of-the-art methods in Section \ref{sec:Experiments}. We conclude this paper in Section \ref{sec:Conclusion}.

\renewcommand{\arraystretch}{1.2}
\begin{table*}[t]
    \footnotesize
    \centering
    \setlength\tabcolsep{3.1pt}
    \caption{\centering The comparison with the existing mixing detection methods.}
    \label{table:method_comparison}
    \begin{tabular}{c|c|c|c|c|c|c|c}
    \hline
    \multirow{3}{*}{\tabincell{c}{Categories}} & \multirow{3}{*}{\tabincell{c}{Refs.}}& \multirow{3}{*}{\tabincell{c}{Methods}}& \multirow{3}{*}{\tabincell{c}{Data Source}}& \multirow{3}{*}{\tabincell{c}{Classifier}} &\multicolumn{3}{c}{\tabincell{c}{Method Characteristics}}  \\ 
    \cline{6-8}
     & & & & &  \multirow{2}{*}{\tabincell{c}{Efficient \\ model training}} & \multirow{2}{*}{\tabincell{c}{Robust to \\noise TX.}} & \multirow{2}{*}{\tabincell{c}{Domain \\portability}} \\
     & & & & & & & \\
    \hline
    \multirow{6}{*}{\tabincell{c}{Service Address \\Identifying}}   
     & MixedSignals \cite{Usenix2023} &   Empirical Study   & Mixing Network Traffic & Expert Judgment & \textcolor{red}\texttimes  & \textcolor{blue}\checkmark & \textcolor{red}\texttimes \\
     & Wu et al. \cite{WWW2021}   &  Heuristic Analysis  & Create Mixing Transaction & Expert Judgment &\textcolor{red}\texttimes& \textcolor{blue}\checkmark & \textcolor{red}\texttimes \\
     & Wu et al. \cite{zhengzibin2022}  &  PU Learning & Web-sourced Labels & LR & \textcolor{blue}\checkmark & \textcolor{red}\texttimes & \textcolor{red}\texttimes \\
     & Xu et al. \cite{IOTJ23}  &  Ensemble Learning & Web-sourced Labels & Random Forest & \textcolor{red}\texttimes & \textcolor{red}\texttimes & \textcolor{red}\texttimes \\
     & Moser et al. \cite{firstmixingstudy}  &  Reverse Engineering & Create Mixing Transaction& Expert Judgment & \textcolor{red}\texttimes & \textcolor{blue}\checkmark & \textcolor{red}\texttimes \\
     & STMD \cite{XiongGang2023}  &  Graph representation Learning & Web-Sourced Labels& MLP & \textcolor{red}\texttimes & \textcolor{red}\texttimes & \textcolor{red}\texttimes \\
    \hline
    \multirow{4}{*}{\tabincell{c}{Input-Output \\Correlation}}  
     & Beres et al. \cite{DAPPS21}& Heuristic Analysis & None& Rule-based Judgment& \textcolor{red}\texttimes & \textcolor{blue}\checkmark & \textcolor{red}\texttimes \\
     & Wang et al. \cite{WWW2023} & Heuristic Analysis  & Side-channel Leakage & Rule-based Judgment & \textcolor{red}\texttimes & \textcolor{red}\texttimes & \textcolor{red}\texttimes \\
     & MixBroker \cite{Mixbroker}& Graph representation Learning& Side-channel Leakage & MLP & \textcolor{red}\texttimes & \textcolor{red}\texttimes & \textcolor{red}\texttimes \\
     \cline{2-8}
     & \textbf{StealthLink} & Graph Transfer Learning & BTC malicious Dataset & MLP&\textcolor{blue}\checkmark &\textcolor{blue}\checkmark & \textcolor{blue}\checkmark \\
    \hline
    \end{tabular}
\end{table*}

\section{Background And Related Work}\label{sec:Related Work}
In this section, we first introduce the background of Tornado Cash, and then we summarize the recent achievements in undermining the anonymity of mixing transactions.

\subsection{Web3.0 and Ethereum}
Web3.0 is a new generation of Internet with the concepts of de-trust, de-intermediation and digital assetisation, with blockchain as the underlying key technology, and digital production and digital consumption as the main economic forms. Web3.0 aims at data sovereign control and value circulation, and through the distributed consensus mechanism, it can completely record the process of value transfer and realise the peer-to-peer transmission of value without the need for specific intermediaries. Through smart contracts, it can form more standard and concise Distributed applications(DApp) to replace the existing Internet application services.

As the largest blockchain platform supporting smart contracts, Ethereum provides a decentralised infrastructure for building Web 3.0 DApps. To guarantee the operation of the decentralised ecosystem, Ethereum sets ETH as the fuel for smart contracts to run, encouraging miners in the network to package and maintain Ethereum transactions. As a result, ETH, as a transaction token in Web3.0, continues to receive attention from both academia and industry.

In the Ethereum blockchain, there are two types of accounts due to the presence of smart contracts: Externally Owned Accounts (EOA) and Contract Accounts (CA). EOAs function similarly to traditional bank accounts in conventional transaction systems, serving as records for transactions between users and their corresponding account balances. In contrast, CAs are internal accounts utilized by users to participate in or invoke various smart contracts. They are responsible for storing information related to smart contracts, such as bytecode and other relevant data. When transactions between EOAs involve smart contracts, they trigger transactions between CAs. Transactions initiated by EOAs are commonly referred to as external transactions, while transactions initiated by Contract Accounts are known as internal transactions. 

\subsection{Tornado Cash}

Tornado Cash, a zk-SNARK-based protocol, operates as a decentralized non-custodial mixing service. Its objective is to enhance transaction privacy by severing the link between source and destination accounts. Through the utilization of smart contracts, Tornado Cash enables the deposit of ETH and other ERC20 tokens from one account and withdrawal from another, seemingly unrelated account. 

Tornado cash Proxy contract acts as a gateway for users to access the TC, and user’s deposit and withdrawal actions are done by interacting with it.

\textbf{Deposit.} Before initiating a deposit, users are required to generate a private $secret$ and a publicly available string $nullifier$ locally. These values are then used to compute the commitment $C$ through a hash function, such that $C = Hash(secret | nullifier)$. Upon initiating a deposit request to the TC contract, users must provide the contract with the commitment $C$ and the specified amount of funds to be deposited, denoted as $N$. The TC contract verifies the availability of the requested ETH amount and subsequently inserts the $C$ as a leaf into the merkel tree list.

\textbf{Withdrawal.} Before initiating a withdrawal, the user, acting as the prover, is required to provide the TC contract, acting as the verifier, with several pieces of information to establish their ownership of the deposited funds. This includes a SNARK proof, the hash value of $nullifier$, the withdrawal account $A$, and the transaction fee $f$. The SNARK proof serves the purpose of demonstrating that the user possesses knowledge of both the merkle path of the $C$ and the preimage of this leaf. The hash value of $nullifier$ is to track spent notes, ensuring that it cannot be reused. Once the TC contract has successfully verified these information, it release the funds to $A$.

In the scenario where the withdrawal account $A$ is a new account without any balance, the TC contract facilitates the transfer of funds to $A$ through a Relayer. The Relayer, acting as an intermediary, deducts a portion of the funds as a transaction fee before transferring the remaining amount to the new account $A$. This transaction fee serves as compensation for the Relayer's services in facilitating the fund transfer.

\subsection{Summary of Existing Studies}
Mixing service have attracted increasing research attention in recent years. In this section, we briefly review the existing Mixing detection into two categories as shown in Table \ref{table:method_comparison}.

\textbf{Service Address Identifying.} This research aims to identify transaction addresses provided by mixing services to track the flow of funds in mixing transactions. Wu et al. \cite{WWW2021} categorized existing mixing techniques into two types, obfuscation and swapping, based on the obfuscation principle, and developed a heuristic approach to identify mixing addresses within the obfuscation mechanism. Fieke et al. \cite{Usenix2023} conducted an empirical study, centering on Bestmixer, utilizing traffic data from mixing servers and publicly available datasets on IP geographic distribution to uncover the underlying principles. Subsequently, researchers began training mixing address classifiers using machine learning models, such as Positive and Unlabeled Learning (PU learning) \cite{zhengzibin2022}, Ensemble Learning \cite{IOTJ23}, and graph representation learning \cite{XiongGang2023}. However, the reliance of machine learning models on expert experience in manual feature design poses challenges in their application to novel mixing mechanisms.

\textbf{Input-Output Correlation.} This research category aims to establish correlations among mixing accounts controlled by the same user through the observation of the user's trading patterns within mixing activities. Two primary technical approaches are employed: heuristic rule design based on expert knowledge \cite{DAPPS21, WWW2023}, and the construction of a graph neural network-based mixing transaction prediction model, utilizing the topological features of mixing interaction graph \cite{Mixbroker}. However, the lack of ground truth sets for mixing transactions presents difficulties in validating the accuracy of the heuristics, while also limiting the precision of the correlation model.

\textbf{Summary.} There are two limitations in the existing methods. On the one hand, existing methods primarily depend on expert-rule design or supervised machine learning techniques \cite{Usenix2023,WWW2021,WWW2023}, which generally require extensive labeled data, with the scarcity of labeled data constraining their effectiveness. On the other hand, the limited available ground datasets are built using heuristic rules, which may incorrectly associate two unrelated addresses due to unintentional actions by users (for example, the association rule based on private transactions \cite{DAPPS21} might erroneously link the sender and recipient of an airdrop transaction), thus introducing noise in the transaction data and further limiting the effectiveness of existing methods.

\section{Problem Definition}\label{sec:Formal definition}
In this paper, we propose a cross-task transfer learning-based model for coin mixing transaction tracing. Specifically, we transfer the knowledge from malicious account detection (source domain $\mathcal{S}$) to the analysis of graph-structured coin mixing transactions (target domain $\mathcal{T}$), enabling effective relational inference under small-sample and noisy conditions in the target domain.

\textbf{Source Domain.} Let $\mathcal{S} = (\mathcal{X}_S, \mathcal{Y}_S, P_S)$ denote the malicious account detection task. Each sample $\mathbf{u}_i \in \mathbb{R}^{d_S}$ in the feature space $\mathcal{X}_S \subseteq \mathbb{R}^{d_S}$ represents a $d_S$-dimensional account feature vector. The label space is defined as $\mathcal{Y}_S = \{0,1\}$, where $y_i = 1$ indicates a malicious account. The source domain contains a large-scale labeled dataset $\mathcal{D}_S^L = \{(\mathbf{u}_i, y_i)\}_{i=1}^m$, where all samples are drawn from the distribution $P_S$, and $m$ denotes the total number of samples in the source domain.

\textbf{Target Domain.} Let $\mathcal{T} = (\mathcal{G}_{T1}, \mathcal{G}_{T2}, \mathcal{Y}_T, P_T)$ denote the target domain, which consists of two heterogeneous coin mixing transaction graphs $\mathcal{G}_{T1}$ and $\mathcal{G}_{T2}$. Specifically, $\mathcal{G}_{T1} = (V_{\alpha1}, V_{\beta}, E_1, X_1)$ represents a transaction subgraph centered around the coin mixing account set $V_{\alpha1} = \{v_{m1}, \dots, v_p\}$, where $V_{\beta} = \{v_{n1}, \dots, v_q\}$ denotes a set of normal account nodes satisfying $V_{\alpha1} \cap V_{\beta} = \varnothing$. The edge set $E_1 \subseteq V_{\alpha1} \times V_{\beta}$ captures the transactional relationships between mixing and normal accounts. The node feature matrix $X_1 \in \mathbb{R}^{p \times d_T}$ encodes $d_T$-dimensional features for mixing accounts.

Similarly, $\mathcal{G}_{T2} = (V_{\alpha2}, V_{\beta}, E_2, X_2)$ denotes a separately constructed transaction subgraph, where the core node set $V_{\alpha2}$ satisfies $V_{\alpha2} \cap (V_{\alpha1} \cup V_{\beta}) = \varnothing$. The label space $\mathcal{Y}_T = \{0,1\}$ is defined over node pairs across the two graphs: for any $u \in V_{\alpha1} \cup V_{\alpha2}$ and $v \in V_{\alpha1} \cup V_{\alpha2}$, $y_{uv} = 1$ indicates that $u$ and $v$ are controlled by the same user. The labeled dataset is given by $\mathcal{D}_T^L = \{(u_k, v_k), y_{uv}\}_{k=1}^n$, where $n$ denotes the total number of labeled node pairs in the target domain, and $m \ll n$.

The target domain is defined as $\mathcal{T} = \{ \mathcal{X}_T, \mathcal{Y}_T \}$, where $\mathcal{X}_T$ denotes the feature space of address pairs involved in coin mixing transactions. Each pair $(\mathbf{v}_i, \mathbf{v}_j) \in \mathbb{R}^{d_T} \times \mathbb{R}^{d_T}$ corresponds to the feature representations of two addresses $\mathbf{v}_i$ and $\mathbf{v}_j$.

The label space $\mathcal{Y}_T = \{0, 1\}$ is a binary set indicating whether the two addresses are associated ($y=1$) or not ($y=0$).

The coin mixing traceability task is formulated as a binary classification problem. Each address pair $(\mathbf{v}_i, \mathbf{v}_j)$ is transformed into a single feature vector $\mathbf{x}_{ij} = \phi(\mathbf{v}_i, \mathbf{v}_j)$, where $\phi: \mathbb{R}^{d_T} \times \mathbb{R}^{d_T} \to \mathbb{R}^{d_T'}$ is a feature fusion function, such as vector concatenation. This transformation produces a dataset $\mathcal{D}_T = \{ (\mathbf{x}_{ij}, y_{ij}) \}$, where $\mathbf{x}_{ij} \in \mathbb{R}^{d_T'}$ is the fused feature vector and $y_{ij} \in \mathcal{Y}_T$ is the corresponding label.

However, due to the high cost of obtaining labeled data, the target domain only contains a limited number of labeled samples, denoted as $\mathcal{D}_T^L \subset \mathcal{D}_T$, with $|\mathcal{D}_T^L| = n$.

\textbf{Design Objective.} The goal is to learn a target domain mapping function $f_T: \mathcal{X}_T \to \mathcal{Y}_T$, where $\mathcal{X}_T = \{ \phi(u,v) \mid u,v \in V_{\alpha1} \cup V_{\alpha2} \}$ denotes the feature space of node pairs. The function $f_T$ should satisfy the following criteria:
\begin{itemize}
    \item \textbf{Few-shot generalization:} When the number of labeled samples is $|\mathcal{D}_T^L| = n \leq 10$, the model should achieve $F_1 \geq 0.80$ on the test set.
    \item \textbf{Noise robustness:} Under label noise with noise rate $\eta \leq 50\%$, the performance degradation should be bounded by $\frac{\Delta_{\text{clean}} - \Delta_{\eta}}{\Delta_{\text{clean}}} \leq 25\%$.
\end{itemize}

\section{Motivation}\label{sec:Transferability Analysis}
Malicious account detection and coin-mixing transaction tracing in blockchain-based cryptocurrencies exhibit significant overlap in their underlying knowledge domains, which provides a solid theoretical foundation for the use of cross-task transfer learning in this chapter. To validate the feasibility of transferring knowledge from the domain of malicious account detection to the domain of coin-mixing transaction association, this section presents an in-depth qualitative and quantitative analysis from the perspectives of domain knowledge and data distribution.

\renewcommand{\arraystretch}{1.2}
\begin{table*}[!t]
\caption{Task Similarity Analysis between Malicious Transaction Detection and Coin-Mixing Traceback}
\label{table:Domain_comparison}
\footnotesize
\renewcommand{\arraystretch}{1.1}
\setlength{\tabcolsep}{4pt}
\begin{center}
\resizebox{\textwidth}{!}{%
\begin{tabular}{c|c|c|c|c|c|c|c|c|c}
\hline
\multirow{2}{*}{Domain} & \multirow{2}{*}{Method} & \multirow{2}{*}{Approach} & \multirow{2}{*}{Task Modeling} & \multirow{2}{*}{Target} & \multicolumn{5}{c}{Feature Attributes} \\
\cline{6-10}
 & & & & & Amount & Time & Frequency & Neighborhood & Fee \\
\hline
\multirow{7}{*}{\makecell{Malicious\\Transaction\\Detection}}
& XBlockFlow \cite{TIFS24ML}   & Empirical Analysis & Binary Classification & Money Laundering & \textcolor{blue}{\checkmark} & \textcolor{blue}{\checkmark} & \textcolor{blue}{\checkmark} & \textcolor{blue}{\checkmark} & \textcolor{red}{\texttimes} \\
& Xiang et al. \cite{2024TIFS-Dataset} & Machine Learning & Multi-class Classification & Malicious Accounts & \textcolor{blue}{\checkmark} & \textcolor{blue}{\checkmark} & \textcolor{blue}{\checkmark} & \textcolor{blue}{\checkmark} & \textcolor{red}{\texttimes} \\
& Bert4ETH \cite{Bert4ETH} & Representation Learning & Binary Classification & Phishing Accounts & \textcolor{blue}{\checkmark} & \textcolor{blue}{\checkmark} & \textcolor{blue}{\checkmark} & \textcolor{blue}{\checkmark} & \textcolor{red}{\texttimes} \\
& ABGRL \cite{TNSE24} & Ensemble Learning & Binary Classification & Phishing Accounts & \textcolor{blue}{\checkmark} & \textcolor{blue}{\checkmark} & \textcolor{blue}{\checkmark} & \textcolor{blue}{\checkmark} & \textcolor{blue}{\checkmark} \\
& Chen et al. \cite{WWW18} & Representation Learning & Binary Classification & Ponzi Schemes & \textcolor{blue}{\checkmark} & \textcolor{blue}{\checkmark} & \textcolor{blue}{\checkmark} & \textcolor{blue}{\checkmark} & \textcolor{blue}{\checkmark} \\
& Jin et al. \cite{JSAC22} & PU Learning & Binary Classification & Arbitrage Accounts & \textcolor{blue}{\checkmark} & \textcolor{blue}{\checkmark} & \textcolor{blue}{\checkmark} & \textcolor{blue}{\checkmark} & \textcolor{red}{\texttimes} \\
& TTAGN \cite{TTAGN} & Representation Learning & Binary Classification & Phishing Accounts & \textcolor{blue}{\checkmark} & \textcolor{blue}{\checkmark} & \textcolor{blue}{\checkmark} & \textcolor{blue}{\checkmark} & \textcolor{red}{\texttimes} \\
\hline
\multirow{6}{*}{\makecell{Coin-Mixing\\Traceback}}
& Wu et al. \cite{WWW2021} & Empirical Analysis & Multi-class Classification & Mixing Transactions & \textcolor{blue}{\checkmark} & \textcolor{blue}{\checkmark} & \textcolor{blue}{\checkmark} & \textcolor{blue}{\checkmark} & \textcolor{red}{\texttimes} \\
& Wang et al. \cite{WWW2023} & Expert Knowledge & Binary Classification & Mixing Transactions & \textcolor{blue}{\checkmark} & \textcolor{blue}{\checkmark} & \textcolor{blue}{\checkmark} & \textcolor{blue}{\checkmark} & \textcolor{blue}{\checkmark} \\
& Wu et al. \cite{zhengzibin2022} & PU Learning & Binary Classification & Mixing Accounts & \textcolor{blue}{\checkmark} & \textcolor{blue}{\checkmark} & \textcolor{blue}{\checkmark} & \textcolor{blue}{\checkmark} & \textcolor{red}{\texttimes} \\
& Xu et al. \cite{IOTJ23} & Ensemble Learning & Binary Classification & Mixing Accounts & \textcolor{blue}{\checkmark} & \textcolor{blue}{\checkmark} & \textcolor{blue}{\checkmark} & \textcolor{blue}{\checkmark} & \textcolor{blue}{\checkmark} \\
& STMD \cite{XiongGang2023} & Representation Learning & Binary Classification & Mixing Accounts & \textcolor{blue}{\checkmark} & \textcolor{blue}{\checkmark} & \textcolor{blue}{\checkmark} & \textcolor{blue}{\checkmark} & \textcolor{blue}{\checkmark} \\
& MixBroker \cite{Mixbroker} & Representation Learning & Binary Classification & Mixing Transactions & \textcolor{blue}{\checkmark} & \textcolor{blue}{\checkmark} & \textcolor{blue}{\checkmark} & \textcolor{blue}{\checkmark} & \textcolor{blue}{\checkmark} \\
\hline
\textbf{Task Similarity} & \multicolumn{9}{p{\dimexpr\textwidth-2\tabcolsep\relax}}{\textbf{All tasks adopt expert-supervised learning methods with similar classification modeling, comparable targets, and highly overlapping feature sets.}} \\
\hline
\end{tabular}
}
\end{center}
\vspace{1mm}

\end{table*}

\subsection{Qualitative Analysis of Domain Transferability}

To qualitatively evaluate the feasibility of transferring knowledge from malicious behavior detection to coin mixing transaction tracing, we conduct a systematic review of recent literature in both domains~\cite{Usenix2023, WWW2021, zhengzibin2022, XiongGang2023, Mixbroker, IOTJ23, 2024TIFS-Dataset, Bert4ETH, TNSE24, WWW18, JSAC22, TTAGN}, as summarized in Table~\ref{table:Domain_comparison}. Our analysis focuses on four key dimensions: analytical methods, task modeling, target objects, and data characteristics.

\textbf{Analytical Methods.} 
Malicious behavior detection primarily relies on two types of approaches. The first involves expert-driven empirical analysis, such as taint analysis~\cite{TIFS24ML}, which traces illicit fund flows to identify suspicious addresses. The second leverages machine learning to automate detection based on features such as transaction statistics and graph structure, using models like MLP~\cite{Bert4ETH} and XGBoost~\cite{JSAC22}. These methodologies are also widely adopted in coin mixing analysis, suggesting a strong methodological overlap between the two domains.

\textbf{Task Modeling.} 
Malicious transaction detection is generally formulated as a binary classification task (benign vs. malicious), aiming to distinguish abnormal behavioral patterns from normal transactions. Given the high anonymity of coin mixing, which is often linked to illicit activities such as money laundering, its behavioral patterns (e.g., high-frequency transactions over short periods~\cite{Maliciousandmixing}) bear strong resemblance to other forms of financial fraud. By modeling transactional similarities between sending and receiving addresses and fusing their features, coin mixing tracing can also be effectively framed as a binary classification task (associated vs. unassociated). This reveals a high degree of alignment in task modeling between the two domains.

\textbf{Target Objects.} 
Existing studies indicate that coin mixing technologies have become critical enablers in the ecosystem of illicit blockchain activities. Anti-money laundering investigations have consistently linked them to darknet markets, ransomware, and other illegal financial flows~\cite{hacker1:online,hacker2:online}. Both tasks focus on blockchain addresses and transaction patterns as core analytical units, leveraging behavioral correlations and fund flow topologies—showing clear structural homogeneity in their analytical targets.

\textbf{Data Characteristics.} 
Due to the intrinsic behavioral similarities between malicious accounts and coin mixing activities, both domains exhibit strong consistency in feature engineering, whether through expert heuristics or automated learning. Key features include transaction timestamps, volumes, frequencies, gas fees, and neighborhood interactions. The high degree of overlap across these multi-dimensional feature spaces underscores the theoretical plausibility of cross-domain knowledge transfer.

\begin{figure}[htbp]
	\small
	\centering
	\includegraphics[width=0.6\linewidth]{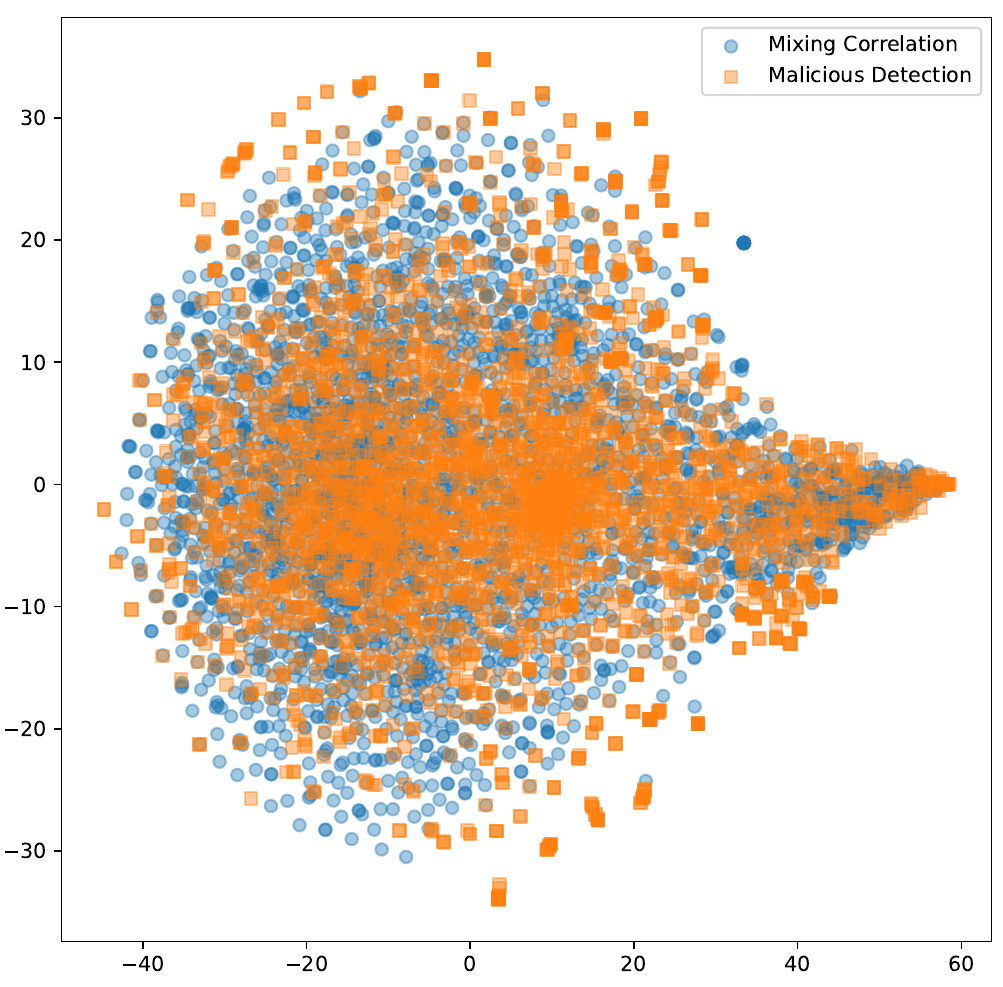}
	\caption{The data distribution of the two domains is highly overlapping.}
	\label{fig:Ini-tsne}
\end{figure}

\subsection{Quantitative Analysis of Domain Transferability}
Based on the qualitative assessment in the previous section, we further perform a quantitative analysis of the transferability from the domain of malicious transaction detection to coin mixing transaction tracing from the perspective of data distribution.

We adopt the Maximum Mean Discrepancy (MMD)~\cite{MMD} metric to evaluate the transferability between the two domains. MMD is a statistical measure used to quantify the difference between two probability distributions. Its core idea is to measure the distance between distributions by comparing the means in a Reproducing Kernel Hilbert Space (RKHS). Notably, a smaller MMD value indicates higher similarity between the two distributions.

For the coin mixing transaction tracing domain, we select the GTD dataset~\cite{Mixbroker} as a representative. This dataset includes 103 pairs of associated input and output addresses from Tornado Cash transactions, each pair described by 46 features including transaction time, amount, gas price, etc. Unassociated samples are constructed by randomly shuffling these address pairs, resulting in 206 address pairs in total.

To avoid bias in MMD computation due to dataset size differences, we choose a reduced version of the BABD-13 dataset~\cite{2024TIFS-Dataset}, denoted as BABD\textsubscript{s}, to represent the malicious behavior detection domain. The BABD\textsubscript{s} dataset is downsized by a factor of 10, containing 54,446 Bitcoin addresses covering 13 behavioral categories, with each address described by 148 features. To ensure consistency in feature dimensionality across datasets, we apply Principal Component Analysis (PCA) to reduce the features of BABD\textsubscript{s} to 46 dimensions.

Let $P$ and $Q$ denote the probability distributions of GTD and BABD\textsubscript{s}, respectively. The MMD is defined as:
\begin{equation}
	\text{MMD}(P, Q) = \left\| \mathbb{E}_{x \sim P}[\phi(x)] - \mathbb{E}_{y \sim Q}[\phi(y)] \right\|_{\mathcal{H}}
\end{equation}

The computed MMD value between the two domains is $2.37 \times 10^{-5}$, which is remarkably small. In practice, MMD values exceeding 0.01 typically indicate significant distributional differences~\cite{MMD}. Hence, this result suggests that the two domains exhibit highly similar feature distributions. We further visualize the data distributions from both domains using the t-distributed Stochastic Neighbor Embedding (t-SNE) algorithm. As shown in Figure~\ref{fig:Ini-tsne}, the data from the two domains show substantial overlap, which corroborates the low MMD value and provides strong quantitative evidence for the similarity in feature distributions between the two tasks.

\section{StealthLink}\label{sec:StealthLink}

\begin{figure*}[htb]
	\centering
	\includegraphics[width=\textwidth]{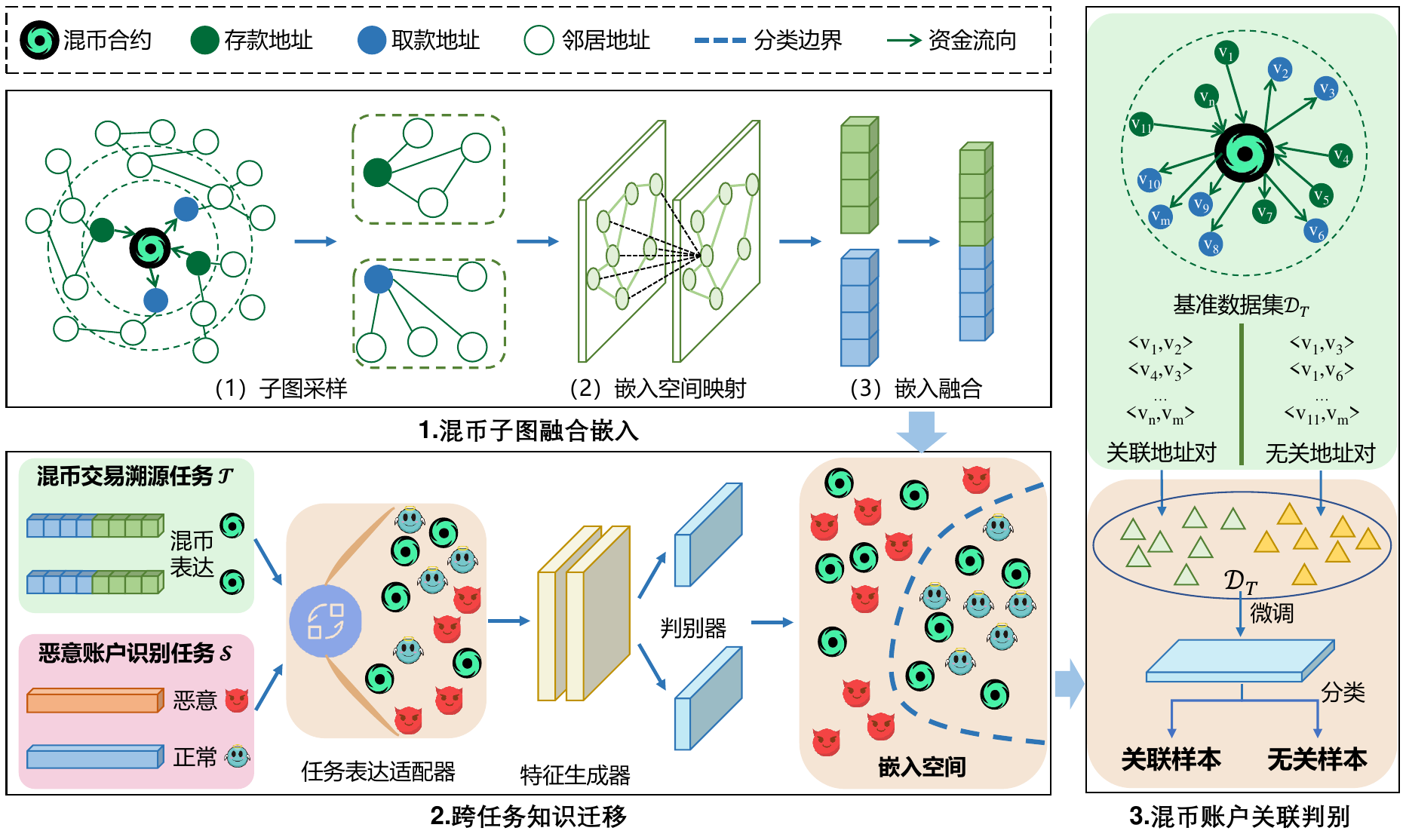}
	\caption{Overview of the StealthLink System}
	\label{fig:StealthLink}
\end{figure*}

In this section, we present the detailed design of \textit{StealthLink}, a method that leverages cross-task knowledge transfer to achieve high-precision tracing of coin mixing transactions under limited labeled samples. The system architecture of StealthLink is illustrated in Figure~\ref{fig:StealthLink}.

\subsection{Overview of the Proposed Approach}\label{subsec:C5-Model}

StealthLink consists of three main components: the Mixed Transaction Subgraph Fusion Embedding Module, the Cross-Task Knowledge Transfer Module, and the Mixed Account Association Discrimination Module.

\textbf{Mixed Transaction Subgraph Fusion Embedding.}  
This module leverages the local topological structure of mixing transaction addresses and graph embedding techniques to capture the latent interaction patterns among addresses involved in coin mixing. It constructs fused sample representations that express the associations between mixed addresses, thereby transforming the complex tracing task into a graph classification problem.

\textbf{Cross-Task Knowledge Transfer.}  
This module introduces a cross-task feature decoupling mechanism to guide the model in learning task-invariant and discriminative shared feature representations. It implicitly aligns the feature space between coin mixing samples and malicious account detection samples, enabling effective cross-task knowledge transfer.

\textbf{Mixed Account Association Discrimination.}  
Based on the pre-trained encoder obtained from the knowledge transfer module, this component employs a few-shot learning strategy to supervise the training of the discriminator. By integrating both transferred and task-specific features, it builds a robust model for tracing coin mixing transactions.

\subsection{Mixed Transaction Subgraph Fusion Embedding}\label{subsec:fusion embedding}

Existing studies on coin mixing transaction association typically analyze the transaction features of individual addresses independently. This approach faces two main limitations. First, the association between mixed addresses is often implicitly embedded in the local behavioral patterns and subgraph structures of both addresses, involving complex nonlinear interactions. Analyzing the transaction attributes of a single address in isolation often fails to capture the combinatorial patterns between address features~\cite{DAPPS21}. Second, due to transaction sparsity (e.g., newly created addresses) and the obfuscation characteristics of mixing techniques (e.g., generating multiple outputs of equal value)~\cite{Mixbroker}, independent address feature analysis often lacks reliable information to determine transaction associations, thereby limiting the effectiveness of standalone embedding methods.

To address these issues, we propose a fusion embedding method tailored for coin mixing subgraphs. First, we use a $k$-hop neighborhood sampling strategy to construct local subgraphs for each of the two addresses involved in a coin mixing transaction, capturing structural and behavioral features in their respective neighborhoods. Second, we employ a graph neural network (GNN) to encode each subgraph, enabling efficient extraction of local representations. Finally, we fuse the two address embeddings through concatenation to generate a joint representation that reflects bidirectional interaction patterns. By integrating the subgraph representations of mixed addresses, the coin mixing association task is formulated as a graph classification problem—determining whether a given joint representation corresponds to an associated address pair.

Specifically, given a coin mixing address pair $v_{\alpha} \in V_{\alpha1}$ and $v_{\beta} \in V_{\alpha2}$, we apply a $k$-hop neighborhood sampling strategy to construct their respective local subgraphs from the blockchain transaction network: $\mathcal{G}_{\alpha} = (V_{\alpha}, E_{\alpha}, X_{\alpha})$ and $\mathcal{G}_{\beta} = (V_{\beta}, E_{\beta}, X_{\beta})$, where $V_{\alpha} = \{v_{\alpha}\} \cup \mathcal{N}_k(v_{\alpha})$ contains the mixed address $v_{\alpha}$ and its $k$-hop neighbors, and $X_{\alpha} \in \mathbb{R}^{|V_{\alpha}| \times d_T}$ is the node feature matrix. Each subgraph is then encoded using a GNN. The embedding of a node $v_i$ at the $l$-th layer is computed as:
\begin{equation}
h_i^{(l+1)} = \sigma\Big( W^{(l)} \cdot \text{AGG}\Big(\{ h_i^{(l)} \} \cup \{ h_j^{(l)} : j \in \mathcal{N}(i) \}\Big) \Big)
\end{equation}
where $W^{(l)}$ is the weight matrix at layer $l$, $\sigma(\cdot)$ is a nonlinear activation function, and $\text{AGG}(\cdot)$ denotes the aggregation function over the neighborhood. After $L$ layers of message passing, we obtain the high-level embeddings $h_\alpha$ and $h_\beta$ for subgraphs $\mathcal{G}_{\alpha}$ and $\mathcal{G}_{\beta}$, respectively. Finally, the two embeddings are concatenated to form a joint representation:
\begin{equation}
\widetilde h = [h_A|h_B] \in \mathbb{R}^{2d_C} 
\end{equation}
where $d_C$ is the embedding dimension for each subgraph. This joint representation comprehensively integrates transaction-level and local structural features of both addresses, enabling subsequent modules to learn domain-invariant feature representations.

\subsection{Cross-task Knowledge Transfer}\label{subsec:transfer learning}

After constructing the joint representation samples for coin-mixing transactions, this section introduces a feature disentanglement mechanism across tasks to guide the model in learning discriminative and shared representations, thereby achieving implicit alignment between coin-mixing transaction samples and malicious account detection samples in the feature space. This enables knowledge transfer across tasks.

Due to significant differences in data representation and feature dimensions between the coin-mixing joint representations and the malicious account detection task—as well as differing focuses on blockchain transaction characteristics and behavior patterns—it is insufficient to directly reuse the representations learned from the malicious account detection task for coin-mixing tracing. To address this, we design a cross-task knowledge transfer module, which consists of two main components: (1) a task representation adapter that maps malicious account detection samples to a feature space compatible with the coin-mixing tracing task; and (2) cross-task invariance learning, which generates domain-aligned and task-discriminative representations in the adapted feature space. The detailed process is as follows:

First, we compute the mean feature vector $\boldsymbol{\mu}_S$ of all encoded samples from the malicious account detection task using encoder $E(\cdot)$ as shown in Eq.~\ref{equ:featureavge}:
\begin{equation}
	\boldsymbol{\mu}_S = \mathbb{E}_{\mathbf{u} \sim P_S}[E(\mathbf{u})]
	\label{equ:featureavge}
\end{equation}
where $E(\cdot): \mathbb{R}^{d_S} \to \mathbb{R}^{d_S}$ denotes the encoder, $\mathbf{u} \in \mathcal{X}_S$ is a sample from the source domain, and $\mathbb{E}_{\mathbf{u} \sim P_S}$ represents the expectation over the source distribution $P_S$. The encoder output $[E(\mathbf{u})] \in \mathbb{R}^{d_S}$ is a high-dimensional feature representation.

Next, a task representation adapter $\mathcal{T}(\cdot)$ is introduced to project the feature vectors from the malicious account detection task into a lower-dimensional space compatible with the coin-mixing tracing task. The transformation is defined in Eq.~\ref{eq:decresedimension}:
\begin{equation}
	\mathcal{T}(\mathbf{u}_i) = U^\top \Big( E(\mathbf{u}_i) - \boldsymbol{\mu}_S \Big)
	\label{eq:decresedimension}
\end{equation}
Here, $U \in \mathbb{R}^{d_S \times d_P}$ is a trainable projection matrix with $d_P = 2d_C$, and $\mathcal{T}(\cdot): \mathbb{R}^{d_S} \to \mathbb{R}^{d_P}$ performs the cross-task feature alignment, ensuring that the adapted feature vectors match the dimensionality of the joint representation $\widetilde h \in \mathbb{R}^{2d_C}$ from the coin-mixing tracing task.

After obtaining the task-aligned representations, we apply a discrepancy-based transfer learning approach to capture invariant features across tasks. This transfer learning process involves training a feature generator \(F(\cdot): \mathbb{R}^{d_P} \to \mathbb{R}^{d_P}\) along with two discriminators $C_1(\cdot)$ and $C_2(\cdot)$. Initially, the feature generator \(F(\cdot)\) is fixed, and the discriminators are trained to maximize the prediction discrepancy on samples from the coin-mixing tracing task. This encourages the discriminators to make diverse predictions in the current feature space. The discrepancy loss is defined as follows:
\begin{equation}
	\mathcal{L}_{\text{dis}} = \mathbb{E}_{\widetilde h \sim P_T} [\| C_1(\widetilde h) - C_2(\widetilde h) \|_1]
	\label{equ:maxdescri}
\end{equation}
where $P_T$ denotes the target distribution (coin-mixing tracing task), and $\|\cdot\|_1$ is the L1 norm.

Next, we fix the discriminators and train the feature generator \(F(\cdot)\) to minimize the discrepancy while also maintaining classification performance on the source domain. The optimization objective is given by Eq.~\ref{equ:mindescri}:
\begin{align}
	\mathcal{L}_{\text{gen}} =\; & \mathbb{E}_{\widetilde h \sim P_T} \left[ \| C_1(\widetilde h) - C_2(\widetilde h) \|_1 \right] \notag \\
	& + \lambda \cdot \mathbb{E}_{(\mathbf{u},y) \sim \mathcal{D}_S^L} 
	\left[ \mathcal{L}_{\text{ce}}\left( C_1\left( F\left(\mathcal{T}(\mathbf{u})\right) \right), y \right) \right]
	\label{equ:mindescri}
\end{align}
where \(\mathcal{L}_{\text{ce}}(p,y)\) denotes the cross-entropy loss, \(\lambda > 0\) is a trade-off parameter balancing the classification loss and the domain alignment loss, and \(\mathcal{D}_S^L = \{(\mathbf{u},y)\}\) is the labeled dataset from the source domain.

Through this learning process of task-invariant features, the feature generator \(F(\cdot)\) is able to produce representations that are both highly discriminative and domain-aligned, enabling effective knowledge transfer from the malicious account detection task to the coin-mixing tracing task. This ultimately enhances the performance of the coin-mixing tracing model in scenarios with limited labeled data.
\subsection{Mixer Account Association Classification}\label{subsec:mix classification}

The mixer account association classification module aims to construct a final discriminator for mixer transaction tracing, based on the task-aligned feature generator trained in the previous module. Building upon the aforementioned task expression adapter, this module introduces a mixer account association classifier $C(\cdot)$, which is fine-tuned using a small labeled set of mixer transaction tracing data $\mathcal{D}_T^L = \{ (\widetilde h_j, y_j) \}_{j=1}^{n}$. During fine-tuning, to ensure the stability of feature representations, the parameters of the feature generator are kept frozen, and only the parameters of the classifier are updated.

For each pair of mixer transaction accounts $(u_j, v_j)$, a fused feature vector $\widetilde h_j \in \mathbb{R}^{2d_C}$ is obtained via the task expression adapter module, and then passed through the classifier $C(\cdot)$ to produce the association prediction. The objective during fine-tuning is to minimize the supervised classification loss, which is computed as shown in Equation~\ref{equ:finetune}:
\begin{equation}
	\mathcal{L}_{ce}^{(T)} = -\frac{1}{n}\sum_{j=1}^{n} \Bigl[ y_j \log C\bigl(\widetilde{h}_j\bigr) + \left(1 - y_j\right) \log \Bigl(1 - C\bigl(\widetilde{h}_j\bigr)\Bigr) \Bigr],
	\label{equ:finetune}
\end{equation}
where $y_j \in \{0,1\}$ indicates whether an association exists between the account pair. Through fine-tuning, the classifier $C(\cdot)$ can effectively leverage the learned cross-task invariant features to accurately determine the association between mixer accounts.

\section{Experiments}\label{sec:Experiments}

In this section, we evaluate the effectiveness of StealthLink on existing datasets using standard metrics, including accuracy, recall, and F1-score. We compare its performance with eight state-of-the-art mixer transaction tracing methods. The experiments demonstrate that StealthLink achieves the following four key capabilities:

(1) In few-shot learning scenarios, StealthLink achieves accurate and robust mixer transaction tracing, outperforming existing models (see Section~\ref{subsec:C5-fewshotlearning});

(2) In scenarios with pseudo-associated noisy samples, StealthLink maintains high accuracy and robustness in tracing, surpassing baseline models (see Section~\ref{subsec:C5-Robustness});

(3) In imbalanced dataset scenarios, StealthLink demonstrates precise and robust tracing capabilities, outperforming current approaches (see Section~\ref{subsec:C5-unbalanced});

(4) Each component of StealthLink contributes to the overall performance in identifying associations between mixer addresses (see Section~\ref{subsec:C5-ablation}).

\subsection{Preliminary}\label{subsec:C5-Preliminary}

\textbf{Experimental Environment.} All experiments were conducted on a Linux-based server equipped with a 16-core Xeon(R) Platinum 8352V processor, 90GB of system memory, and an NVIDIA GeForce RTX 4090 GPU (24GB VRAM) with driver version 560.35.03. The software environment includes Python 3.7 and PyTorch 1.13.1, with CUDA Toolkit 11.7.

\textbf{Datasets.} This section involves three types of datasets: the Bitcoin malicious account detection dataset, the mixer transaction dataset, and the mixer benchmark dataset.

\begin{itemize}
    \item \textbf{Bitcoin Malicious Account Detection Dataset.} This dataset is constructed based on the BABD-13 dataset~\cite{2024TIFS-Dataset}, which contains fine-grained labels for six types of malicious accounts (phishing, gambling, darknet markets, blacklisted addresses, money laundering, and Ponzi schemes) and seven types of benign accounts. In our experimental design, the six malicious account types are merged into a single malicious class, forming a binary classification task together with the benign accounts. The final dataset contains 41,662 benign accounts and 12,842 malicious accounts, totaling 54,504 samples.
    
    \item \textbf{Tornado Cash Transaction Dataset.} We used the Etherscan API to crawl all mixing transaction data from the launch of Tornado Cash up to March 31, 2022. This resulted in 30,823 deposit addresses and 44,814 withdrawal addresses. Any combination of a deposit and a withdrawal address is treated as an unlabeled mixer transaction sample.
    
    \item \textbf{Mixing Benchmark Dataset.} This dataset aggregates labeled Ethereum address pairs from the studies~\cite{Mixbroker, DAPPS21}, consisting of a total of 291 associated address pairs.
\end{itemize}

\textbf{Baseline Methods.} To comprehensively evaluate the performance of StealthLink, we compare it with eight state-of-the-art address association methods for mixer transactions. All baselines were fine-tuned to ensure optimal performance on the evaluation datasets.

\begin{itemize}
    \item \textbf{Gas Fingerprinting (GF)}~\cite{WWW2023}: A heuristic method that matches transactions by identifying cases where the last 9 digits of the gas price are identical in both sending and receiving transactions.

    \item \textbf{Cross Contract Correlation (CC)}~\cite{WWW2023}: Another heuristic method that links deposit and withdrawal addresses across multiple mixer pools, leveraging the fixed denomination feature of Tornado Cash and user behaviors involving multiple deposits.

    \item \textbf{DeepWalk}~\cite{perozzi2014deepwalk}: Generates random walks in the transaction subgraph to produce sequences of nodes, which are then embedded into a low-dimensional vector space. Similarity between embeddings is used to infer address associations.

    \item \textbf{Node2Vec}~\cite{grover2016node2vec}: Extends DeepWalk with two hyperparameters that balance breadth-first and depth-first sampling, achieving better trade-offs between local and global structural features.

    \item \textbf{GAT}~\cite{GAT}: Utilizes self-attention mechanisms to assign different weights to neighboring nodes in the transaction subgraph of a mixer address, generating low-dimensional embeddings. Address similarity is then assessed via these embeddings.

    \item \textbf{GIN}~\cite{GIN}: Aggregates neighborhood information in the transaction subgraph using multi-layer perceptrons (MLPs), capturing subtle subgraph structural differences through low-dimensional embeddings.

    \item \textbf{GraphSAGE}~\cite{GraphSAGE}: Samples and aggregates neighborhood node information in the transaction subgraph to produce embeddings, which are then used to evaluate address similarity.

    \item \textbf{MixBroker}~\cite{Mixbroker}: Models Tornado Cash address relationships via interaction graphs, and inputs statistical features into a GNN-based classifier to analyze address associations.
\end{itemize}

\textbf{Parameter Settings.} For StealthLink, we adopt a two-stage training strategy. In the pretraining stage, we use a Transformer-based encoder consisting of 3 stacked layers with parameters: \texttt{d\_model} = 92, \texttt{nhead} = 4. The output features are projected via a head composed of linear layers, batch normalization (BN), and ReLU activation. The pretraining uses SGD optimizer with a learning rate of \(1 \times 10^{-4}\), momentum of 0.9, and weight decay of 0.0005. During the fine-tuning stage, the encoder's output embeddings are frozen and passed to a multi-layer perceptron (MLP) classifier. The MLP consists of four hidden layers with 1024 units each, an input dimension of 92, an output dimension of 2, ReLU activation for hidden layers, and a Sigmoid activation at the output layer.

For DeepWalk and Node2Vec, the embedding dimension is set to 43, with walk length, number of walks, and context window size all set to 5. Node2Vec's return and in-out parameters are both set to 0.75, using the Skip-gram model (\texttt{sg}=1) with 4 worker threads.

For GAT, the input feature dimension is set to 31, hidden layer dimension to 43, number of classes to 30, and number of attention heads to 4.  

For GIN, the model consists of 2 layers, each with a 3-layer MLP. Input, hidden, and output dimensions are 31, 256, and 43 respectively. Both \texttt{graph\_pooling\_type} and \texttt{neighbor\_pooling\_type} are set to "mean", and \texttt{final\_dropout} is 0.01.  

For GraphSAGE, the number of input channels is 31, hidden channels is 128, number of layers is 2, and output channels is 43.

For MixBroker, the model is configured with \texttt{GNN\_NET}, where the GNN comprises two SAGEConv layers: the first maps input features to a 32-dimensional hidden space, and the second maps to a 16-dimensional output. The optimizer is Adam, with a learning rate of 0.01.

\subsection{Few-shot Learning Evaluation}\label{subsec:C5-fewshotlearning}
This section presents a systematic evaluation of StealthLink's performance under few-shot learning scenarios. Specifically, we conduct a quantitative analysis of its capability in link prediction for mixing transactions when the training set consists of only 1, 3, 5, or 10 samples. To mitigate the impact of sample selection randomness, we generate 10 independent training sets for each sample size via random resampling. The final results are reported as the mean ± standard deviation across the 10 trials. 

To establish performance baselines, we also evaluate the model trained on the full labeled dataset of mixing transactions. It is important to note that rule-based methods (GF and CC), which do not involve parameter training, are evaluated only under the full-data setting. The quantitative results under different experimental configurations are summarized in Table~\ref{tab:C5-FewshotLe}.

\begin{table*}[htbp]
\centering
\caption{Performance Comparison Under Few-Shot Scenarios}
\label{tab:C5-FewshotLe}
\resizebox{\textwidth}{!}{
\begin{tabular}{ccccccccccc}
\toprule
\multirow{2}*{N} & \multirow{2}*{Metric} & \multicolumn{9}{c}{Method} \\
\cmidrule(lr){3-11}
& & GF & CC & DeepWalk & Node2vec & GAT & GIN & GraphSAGE & MixBroker & StealthLink \\
\midrule

\multirow{3}*{1}
& Accuracy & -- & -- & 0.4309$\pm$0.0264 & 0.4193$\pm$0.0353 & 0.4832$\pm$0.1163 & 0.7155$\pm$0.0904 & 0.4865$\pm$0.0326 & 0.7851$\pm$0.1138 & 0.7879$\pm$0.1288 \\
& Recall    & -- & -- & 0.3979$\pm$0.0627 & 0.4382$\pm$0.1689 & 0.4774$\pm$0.3142 & 0.6776$\pm$0.1404 & 0.6215$\pm$0.2460 & 0.7277$\pm$0.2409 & 0.8324$\pm$0.0484 \\
& F1-score  & -- & -- & 0.4138$\pm$0.0325 & 0.4285$\pm$0.0689 & 0.4803$\pm$0.2486 & 0.6788$\pm$0.0421 & 0.5458$\pm$0.0980 & 0.7554$\pm$0.1453 & \textbf{0.8096$\pm$0.0678} \\
\midrule

\multirow{3}*{3}
& Accuracy & -- & -- & 0.4295$\pm$0.0332 & 0.4187$\pm$0.0447 & 0.7264$\pm$0.0840 & 0.8642$\pm$0.0751 & 0.5491$\pm$0.0365 & 0.7881$\pm$0.1179 & 0.9429$\pm$0.0199 \\
& Recall    & -- & -- & 0.4062$\pm$0.0670 & 0.4579$\pm$0.1420 & 0.7079$\pm$0.0801 & 0.7048$\pm$0.0872 & 0.5140$\pm$0.1260 & 0.7515$\pm$0.1732 & 0.9738$\pm$0.0167 \\
& F1-score  & -- & -- & 0.4175$\pm$0.0270 & 0.4375$\pm$0.0772 & 0.7170$\pm$0.0451 & 0.7713$\pm$0.0562 & 0.5310$\pm$0.0633 & 0.7694$\pm$0.0959 & \textbf{0.9580$\pm$0.0118} \\
\midrule

\multirow{3}*{5}
& Accuracy & -- & -- & 0.4591$\pm$0.0339 & 0.4490$\pm$0.0186 & 0.7021$\pm$0.0560 & 0.8563$\pm$0.0864 & 0.5830$\pm$0.0320 & 0.7986$\pm$0.2921 & 0.9657$\pm$0.0085 \\
& Recall    & -- & -- & 0.4706$\pm$0.0986 & 0.4632$\pm$0.0935 & 0.7395$\pm$0.0885 & 0.7790$\pm$0.0873 & 0.5227$\pm$0.1263 & 0.8106$\pm$0.2857 & 0.9660$\pm$0.0104 \\
& F1-score  & -- & -- & 0.4648$\pm$0.0375 & 0.4560$\pm$0.0543 & 0.7203$\pm$0.0217 & 0.8084$\pm$0.0475 & 0.5512$\pm$0.0647 & 0.8026$\pm$0.2877 & \textbf{0.9657$\pm$0.0044} \\
\midrule

\multirow{3}*{10}
& Accuracy & -- & -- & 0.4564$\pm$0.0300 & 0.4654$\pm$0.0148 & 0.7323$\pm$0.0448 & 0.8313$\pm$0.0257 & 0.6452$\pm$0.4437 & 0.8988$\pm$0.0783 & 0.9741$\pm$0.0031 \\
& Recall    & -- & -- & 0.3822$\pm$0.0557 & 0.4266$\pm$0.0858 & 0.7597$\pm$0.0623 & 0.7552$\pm$0.0705 & 0.5982$\pm$0.0739 & 0.7409$\pm$0.1027 & 0.9662$\pm$0.0036 \\
& F1-score  & -- & -- & 0.4161$\pm$0.0304 & 0.4452$\pm$0.0445 & 0.7458$\pm$0.0202 & 0.7916$\pm$0.0422 & 0.6208$\pm$0.0468 & 0.8122$\pm$0.1076 & \textbf{0.9698$\pm$0.0015} \\
\midrule

\multirow{3}*{ALL}
& Accuracy & 0.9291 & 1.0000 & 0.4979$\pm$0.1478 & 0.4939$\pm$0.1698 & 0.8808$\pm$0.1491 & 0.8512$\pm$0.0274 & 0.8006$\pm$0.0508 & 0.8781$\pm$0.1040 & 0.9832$\pm$0.0082 \\
& Recall    & 0.1265 & 0.0681 & 0.5562$\pm$0.2549 & 0.6140$\pm$0.2787 & 0.6971$\pm$0.3785 & 0.7982$\pm$0.0186 & 0.8848$\pm$0.0484 & 0.8473$\pm$0.1250 & 0.9913$\pm$0.0057 \\
& F1-score  & 0.2226 & 0.1275 & 0.5255$\pm$0.1912 & 0.5474$\pm$0.2117 & 0.7782$\pm$0.3380 & 0.8239$\pm$0.0130 & 0.8403$\pm$0.0274 & 0.8548$\pm$0.0740 & \textbf{0.9872$\pm$0.0103} \\
\bottomrule
\end{tabular}
}
\end{table*}
As shown in Table~\ref{tab:C5-FewshotLe}, StealthLink demonstrates consistently strong discriminative performance across all few-shot scenarios. This superior performance can be attributed to the combination of cross-task invariant feature learning and the task representation adapter module, which enables the model to effectively activate large-scale knowledge of malicious account detection even with very limited mixing transaction samples, thereby significantly improving accuracy under low-resource conditions. Remarkably, even with as few as \(N=3\) labeled samples, StealthLink achieves an F1 score of 0.9580, which significantly outperforms all other baselines trained on the full dataset. For example, MixBroker only achieves an F1 score of 0.8122 under full supervision, indicating the superior capability of the proposed method in few-shot learning scenarios for mixing transaction tracing.

In contrast, two heuristic-based methods, GF and CC, exhibit poor performance. Under full supervision, GF and CC achieve F1 scores of only 0.2226 and 0.1275, respectively. This underperformance is due to the heuristic methods relying solely on local and isolated features for address linkage in mixing transactions, making them incapable of adapting to the complex and dynamic structure of mixing transaction networks. Consequently, they struggle to extract deeper feature representations, resulting in limited scalability and effectiveness in large-scale mixing transaction tracing tasks.

\subsection{Robustness Evaluation under Noisy Labels}\label{subsec:C5-Robustness}

This section systematically evaluates the model's robustness under label noise. Given that existing annotated datasets are constructed based on heuristic rules~\cite{DAPPS21,Mixbroker}, they inherently contain mislabeled address association pairs in mixing transactions. To address this, we construct a controlled noise injection environment by introducing 5\%--50\% of false association samples into the training set. This allows us to quantitatively analyze the performance stability of StealthLink in the task of mixing address association under noisy conditions. A ten-fold cross-validation strategy is employed, and the model’s performance under noise interference is reported in the form of mean ± standard deviation. Table~\ref{tab:C5-robutness} presents the comparative results of quantitative evaluations under different levels of noise.

\begin{table*}[htbp]
	\centering
	\caption{Performance Comparison under Different Noise Ratios}
	\label{tab:C5-robutness}
	\resizebox{\textwidth}{!}{
	\begin{tabular}{ccccccccc}
		\toprule
		\multirow{2}*{Noise Ratio} & \multirow{2}*{Metric} & \multicolumn{7}{c}{Method} \\
		\cmidrule(lr){3-9}
		& & DeepWalk & Node2vec & GAT & GIN & GraphSAGE & MixBroker & StealthLink \\
		\midrule
		
		\multirow{3}*{5\%}
		& Accuracy & 0.4538$\pm$0.0293 & 0.4513$\pm$0.0224 & 0.8541$\pm$0.0250 & 0.8468$\pm$0.0370 & 0.6501$\pm$0.0315 & 0.8822$\pm$0.0462 & \textbf{0.9665$\pm$0.0103} \\
		& Recall & 0.4717$\pm$0.0919 & 0.5014$\pm$0.1126 & 0.6891$\pm$0.0439 & 0.7693$\pm$0.0849 & 0.7072$\pm$0.1322 & 0.7351$\pm$0.1000 & \textbf{0.9321$\pm$0.0951} \\
		& F1 Score & 0.4626$\pm$0.0493 & 0.4750$\pm$0.0582 & 0.7627$\pm$0.0144 & 0.8062$\pm$0.0437 & 0.6774$\pm$0.0467 & 0.8020$\pm$0.0615 & \textbf{0.9490$\pm$0.0526} \\
		\midrule
		\multirow{3}*{10\%}
		& Accuracy & 0.4249$\pm$0.0323 & 0.4594$\pm$0.0290 & 0.8734$\pm$0.0264 & 0.8343$\pm$0.0501 & 0.5909$\pm$0.0279 & 0.8516$\pm$0.0167 & \textbf{0.9165$\pm$0.0103} \\
		& Recall & 0.4837$\pm$0.1656 & 0.4434$\pm$0.1244 & 0.6653$\pm$0.0439 & 0.7120$\pm$0.1202 & 0.6508$\pm$0.1575 & 0.6706$\pm$0.0334 & \textbf{0.9017$\pm$0.0015} \\
		& F1 Score & 0.4524$\pm$0.0941 & 0.4512$\pm$0.0706 & 0.7553$\pm$0.0196 & 0.7684$\pm$0.0662 & 0.6194$\pm$0.0663 & 0.7504$\pm$0.0272 & \textbf{0.9090$\pm$0.0052} \\
		\midrule
		\multirow{3}*{20\%}
		& Accuracy & 0.4071$\pm$0.0316 & 0.4371$\pm$0.0263 & 0.8113$\pm$0.0331 & 0.8376$\pm$0.0506 & 0.5373$\pm$0.0384 & 0.7902$\pm$0.0042 & \textbf{0.8896$\pm$0.0158} \\
		& Recall & 0.4074$\pm$0.1846 & 0.4477$\pm$0.1445 & 0.6842$\pm$0.0771 & 0.6764$\pm$0.1366 & 0.6375$\pm$0.2147 & 0.6154$\pm$0.0261 & \textbf{0.8721$\pm$0.0137} \\
		& F1 Score & 0.4073$\pm$0.1095 & 0.4424$\pm$0.0834 & 0.7424$\pm$0.0225 & 0.7486$\pm$0.0729 & 0.5831$\pm$0.1079 & 0.6920$\pm$0.0266 & \textbf{0.8808$\pm$0.0080} \\
		\midrule
		\multirow{3}*{30\%}
		& Accuracy & 0.4263$\pm$0.0281 & 0.3791$\pm$0.0346 & 0.7686$\pm$0.0482 & 0.8048$\pm$0.0500 & 0.4628$\pm$0.0333 & 0.7633$\pm$0.0051 & \textbf{0.8331$\pm$0.0137} \\
		& Recall & 0.3989$\pm$0.1533 & 0.4117$\pm$0.2087 & 0.6283$\pm$0.1033 & 0.6366$\pm$0.1407 & 0.5925$\pm$0.3000 & 0.5876$\pm$0.0083 & \textbf{0.8217$\pm$0.0108} \\
		& F1 Score & 0.4122$\pm$0.0913 & 0.3948$\pm$0.1396 & 0.6914$\pm$0.0465 & 0.7108$\pm$0.0719 & 0.5198$\pm$0.2005 & 0.6643$\pm$0.0083 & \textbf{0.8274$\pm$0.0128} \\
		\midrule
		\multirow{3}*{50\%}
		& Accuracy & 0.3114$\pm$0.0457 & 0.3644$\pm$0.0300 & 0.4297$\pm$0.0897 & 0.6177$\pm$0.0386 & 0.4528$\pm$0.0342 & 0.6842$\pm$0.0290 & \textbf{0.8001$\pm$0.1306} \\
		& Recall & 0.3325$\pm$0.2370 & 0.3351$\pm$0.1647 & 0.4480$\pm$0.2148 & 0.5826$\pm$0.2431 & 0.4735$\pm$0.2015 & 0.4818$\pm$0.0580 & \textbf{0.7285$\pm$0.1412} \\
		& F1 Score & 0.3217$\pm$0.1718 & 0.3491$\pm$0.1325 & 0.4387$\pm$0.1533 & 0.5998$\pm$0.1614 & 0.4629$\pm$0.1014 & 0.5654$\pm$0.0533 & \textbf{0.7629$\pm$0.0771} \\
		\bottomrule
	\end{tabular}
	}
\end{table*}
As shown in Table~\ref{tab:C5-robutness}, StealthLink demonstrates a significant robustness advantage on datasets contaminated with pseudo-associated noisy labels. Under all noise ratios, StealthLink consistently achieves higher F1-scores compared to other methods, outperforming the second-best approach, MixBroker, by 15\%–20\%. This improvement is attributed to the model's ability to learn general representations through cross-task invariant feature learning, combined with a dynamic adversarial learning framework consisting of a feature generator and dual discriminators. This architecture effectively filters out noise interference, thereby significantly enhancing the model’s discriminative capability under high-noise conditions.

\begin{figure}[t]
    \centering
    \includegraphics[width=\columnwidth]{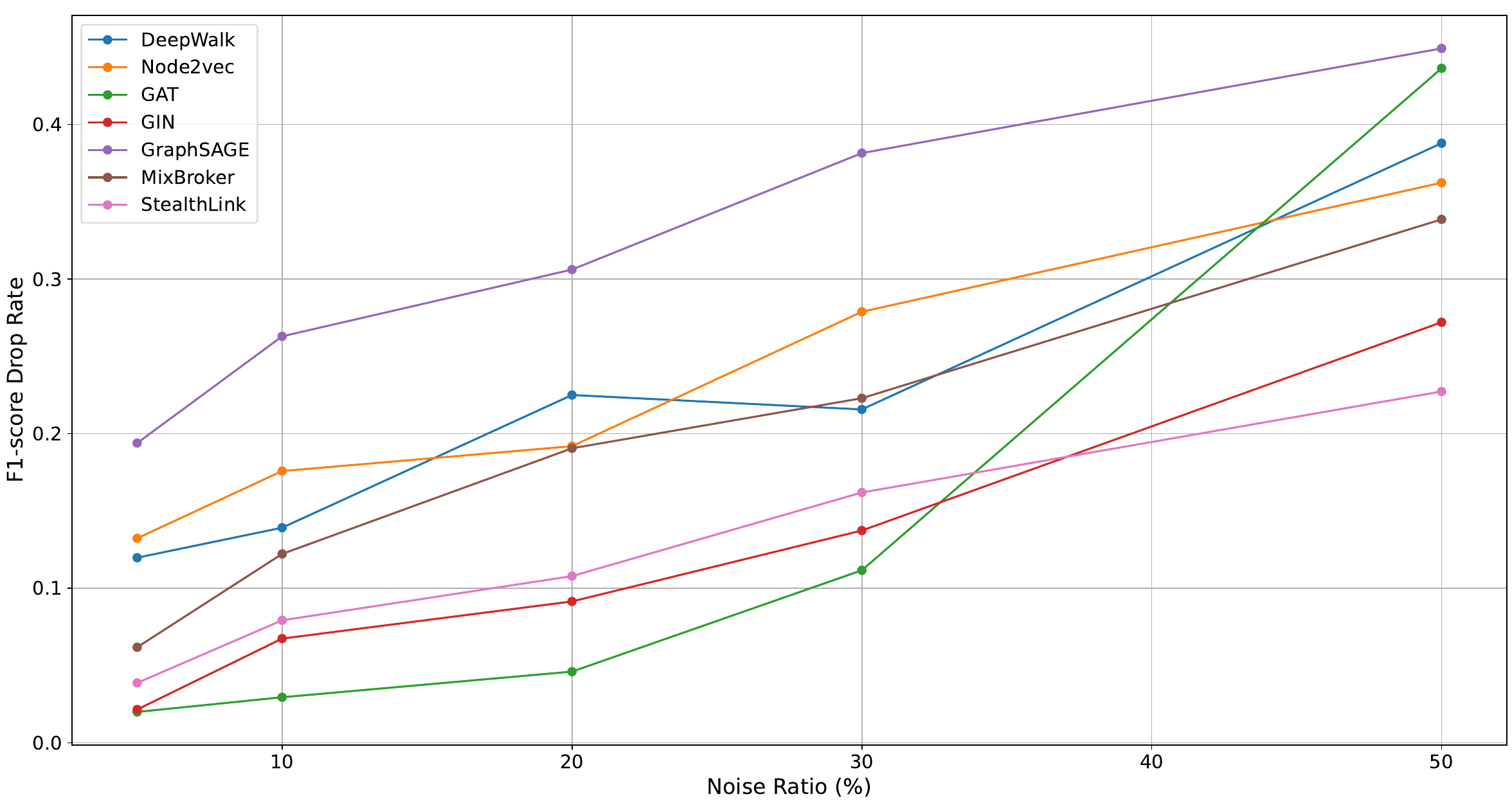}
    \caption{Relationship between label noise rate and performance degradation rate}
    \label{fig:drop-rate}
\end{figure}

To further evaluate the noise resistance of StealthLink, we analyze the relationship between label noise rates and performance degradation rates, as shown in Fig.~\ref{fig:drop-rate}. It can be observed that StealthLink exhibits a markedly lower slope in its performance degradation curve compared to baseline methods. Specifically, its F1-score degradation rate remains below 0.25, indicating stronger robustness to label noise. These results confirm the stability and robustness of StealthLink in noisy environments.

\subsection{Evaluation on Imbalanced Datasets}\label{subsec:C5-unbalanced}
This section further evaluates the performance of StealthLink in the task of association discrimination under imbalanced positive and negative sample training scenarios. To control the influence of sample randomness, random sampling is used, and the mean is calculated based on 10-fold cross-validation. The classification results are tested by constructing four different positive-to-negative sample ratios: 1:5, 1:10, 1:15, and 1:25. The experimental results are shown in Table \ref{tab:C5-IMBALANCE}.

\begin{table*}[htbp]
    \centering
    \caption{Performance Comparison on Imbalanced Datasets}
    \label{tab:C5-IMBALANCE}
    \resizebox{\textwidth}{!}{
    \begin{tabular}{ccccccccc}
        \toprule
        \multirow{2}*{Positive-to-Negative Ratio} & \multirow{2}*{Evaluation Metric} & \multicolumn{7}{c}{Methods} \\
        \cmidrule(lr){3-9}
        & & GIN & GAT & GraphSAGE & DeepWalk & Node2Vec & MixBroker & StealthLink \\
        \midrule

        \multirow{3}*{1:5}
        & Accuracy & 0.8361$\pm$0.0024 & 0.8351$\pm$0.2746 & 0.7979$\pm$0.0193 & 0.3087$\pm$0.1068 & 0.7063$\pm$0.3529 & 0.8961$\pm$0.0695 & 0.9741$\pm$0.0031 \\
        & Recall & 0.7569$\pm$0.0945 & 0.4761$\pm$0.2179 & 0.8163$\pm$0.1338 & 0.1913$\pm$0.0548 & 0.3326$\pm$0.1594 & 0.7460$\pm$0.0960 & 0.9162$\pm$0.0036 \\
        & F1 Score & 0.7944$\pm$0.0567 & 0.6065$\pm$0.2547 & 0.8070$\pm$0.0844 & 0.2362$\pm$0.0718 & 0.4522$\pm$0.2190 & 0.8094$\pm$0.0538 & \textbf{0.9443$\pm$0.0015} \\
        \midrule
        \multirow{3}*{1:10}
        & Accuracy & 0.7964$\pm$0.0011 & 0.7706$\pm$0.3554 & 0.7858$\pm$0.0125 & 0.3977$\pm$0.2364 & 0.7072$\pm$0.3676 & 0.9213$\pm$0.0818 & 0.9257$\pm$0.0085 \\
        & Recall & 0.7139$\pm$0.0599 & 0.4334$\pm$0.2700 & 0.6903$\pm$0.2167 & 0.1183$\pm$0.0642 & 0.2677$\pm$0.1488 & 0.7181$\pm$0.0742 & 0.9260$\pm$0.0104 \\
        & F1 Score & 0.7530$\pm$0.0335 & 0.5548$\pm$0.3293 & 0.7367$\pm$0.1676 & 0.1824$\pm$0.0988 & 0.3884$\pm$0.2116 & 0.8033$\pm$0.0530 & \textbf{0.9258$\pm$0.0044} \\
        \midrule
        \multirow{3}*{1:15}
        & Accuracy & 0.7164$\pm$0.2988 & 0.7801$\pm$0.3413 & 0.7323$\pm$0.0131 & 0.4910$\pm$0.2956 & 0.7143$\pm$0.3649 & 0.8924$\pm$0.0658 & 0.9129$\pm$0.0199 \\
        & Recall & 0.6817$\pm$0.2911 & 0.4368$\pm$0.2668 & 0.6537$\pm$0.2032 & 0.1069$\pm$0.0610 & 0.2528$\pm$0.1498 & 0.6942$\pm$0.0747 & 0.9038$\pm$0.0167 \\
        & F1 Score & 0.6987$\pm$0.2947 & 0.5600$\pm$0.3237 & 0.6908$\pm$0.1709 & 0.1756$\pm$0.0950 & 0.3734$\pm$0.2135 & 0.7789$\pm$0.0602 & \textbf{0.9085$\pm$0.0118} \\
        \midrule
        \multirow{3}*{1:25}
        & Accuracy & 0.6972$\pm$0.2990 & 0.8048$\pm$0.3325 & 0.7302$\pm$0.0093 & 0.5946$\pm$0.3688 & 0.7764$\pm$0.3447 & 0.8826$\pm$0.0489 & 0.7879$\pm$0.1288 \\
        & Recall & 0.6637$\pm$0.2884 & 0.4315$\pm$0.2681 & 0.5250$\pm$0.1640 & 0.1099$\pm$0.0670 & 0.2352$\pm$0.1452 & 0.6768$\pm$0.0715 & 0.8324$\pm$0.0484 \\
        & F1 Score & 0.6801$\pm$0.2934 & 0.5618$\pm$0.3272 & 0.6109$\pm$0.1504 & 0.1856$\pm$0.1129 & 0.3611$\pm$0.2083 & 0.7634$\pm$0.0484 & \textbf{0.8096$\pm$0.0678} \\
        \bottomrule
    \end{tabular}
    }
\end{table*}

As shown in Table \ref{tab:C5-IMBALANCE}, StealthLink maintains excellent discriminatory performance even under extreme class imbalance conditions. Specifically, as the positive-to-negative sample ratio increases from 1:5 to 1:25, its F1-score exhibits a stepwise decline (0.9443 $\rightarrow$ 0.8096), but it still consistently outperforms the baseline models. Meanwhile, the second-best benchmark model, MixBroker, experiences a larger F1-score drop of 9.8\% (0.8094 $\rightarrow$ 0.7634) under the same test conditions, resulting in a significant difference when compared to StealthLink. This phenomenon can be attributed to the synergistic effect of StealthLink's cross-task invariance feature learning and adversarial discriminator difference minimization, which effectively preserves and aligns key discriminative features even in highly imbalanced conditions, thus balancing high precision and recall. This enables StealthLink to robustly perform coin-mixing transaction link prediction, even when minority class samples are extremely scarce.

\subsection{Ablation Study}\label{subsec:C5-ablation}
This section quantitatively analyzes the impact of different components on the performance of StealthLink, including the task expression adapter, feature generator, and transfer learning paradigm in the cross-task knowledge transfer module, as well as the classifier in the coin-mixing account linkage module. Since the framework design focuses on coin-mixing entity linkage under small-sample training scenarios, this experiment primarily evaluates the impact of different components on StealthLink's performance in such small-sample learning contexts.

\textbf{Task Expression Adapter Evaluation.} The task expression adapter aligns the feature space of malicious account detection and coin-mixing transaction tracing tasks through feature mapping strategies. This section evaluates the impact of three common feature mapping strategies on StealthLink's performance: dimensionality expansion, PCA dimensionality reduction, and contribution pruning. Dimensionality expansion projects the low-dimensional source domain representation to the target domain's dimension using a Multi-Layer Perceptron (MLP); PCA dimensionality reduction compresses the high-dimensional source domain features to the target dimension using Principal Component Analysis (PCA); contribution pruning iteratively removes low-importance features based on feature contribution ranking until the target dimension is reached. We evaluate the model performance under each strategy using ten-fold cross-validation, with the results presented in Table \ref{tab:C5-dim_performance}.

\begin{table*}[htbp]
	\centering
	\caption{Performance Comparison of StealthLink with Different Feature Mapping Strategies}
	\label{tab:C5-dim_performance}
	\begin{tabular}{ccccc}
		\toprule
		\multirow{2}*{N} & \multirow{2}*{Evaluation Metric} & \multicolumn{3}{c}{Alignment Method} \\
		\cmidrule(lr){3-5}
		& & Dimensionality Expansion & PCA & Contribution Pruning \\
		\midrule
		\multirow{3}*{1} 
		& Accuracy & 0.6672$\pm$0.2501 & 0.7879$\pm$0.1288 & 0.5969$\pm$0.1733 \\
		& Recall & 0.6759$\pm$0.1372 & 0.8324$\pm$0.0484 & 0.7655$\pm$0.2329 \\
		& F1 Score & 0.6715$\pm$0.1004 & \textbf{0.8096$\pm$0.0678} & 0.6708$\pm$0.1815 \\
		\midrule
		\multirow{3}*{3}
		& Accuracy & 0.8915$\pm$0.1396 & 0.9429$\pm$0.0199 & 0.7179$\pm$0.0831 \\
		& Recall & 0.6830$\pm$0.1024 & 0.9738$\pm$0.0167 & 0.8305$\pm$0.0793 \\
		& F1 Score & 0.7735$\pm$0.0402 & \textbf{0.9580$\pm$0.0118} & 0.7701$\pm$0.0395 \\
		\midrule
		\multirow{3}*{5}
		& Accuracy & 0.8205$\pm$0.1732 & 0.9657$\pm$0.0085 & 0.7956$\pm$0.1139 \\
		& Recall & 0.6415$\pm$0.1846 & 0.9660$\pm$0.0104 & 0.8418$\pm$0.1471 \\
		& F1 Score & 0.7314$\pm$0.0279 & \textbf{0.9657$\pm$0.0044} & 0.8108$\pm$0.1062 \\
		\midrule
		\multirow{3}*{10}
		& Accuracy & 0.9153$\pm$0.0874 & 0.9741$\pm$0.0031 & 0.7799$\pm$0.1124 \\
		& Recall & 0.6718$\pm$0.1024 & 0.9662$\pm$0.0036 & 0.8635$\pm$0.1190 \\
		& F1 Score & 0.7749$\pm$0.0615 & \textbf{0.9698$\pm$0.0015} & 0.8196$\pm$0.0877 \\
		\bottomrule
	\end{tabular}
\end{table*}

As shown in Table \ref{tab:C5-dim_performance}, the PCA dimensionality reduction strategy achieves the highest F1 score across all data size settings, indicating that the PCA reduction strategy can maintain the model's coin-mixing transaction address association ability in scenarios with varying label sparsity. This is because PCA effectively extracts the core feature distribution shared by the malicious account detection task and the coin-mixing transaction tracing task by retaining the principal components with the largest global variance. It reduces noise dimensions while preserving the integrity of key features. In contrast, the dimensionality expansion strategy performs poorly in scenarios with a small number of label samples, likely because its reliance on the MLP trained with limited labeled data leads to overfitting or difficulties in capturing the feature differences between the two tasks. Meanwhile, contribution pruning may lose some subtle but discriminative feature dimensions when pruning low-contribution features, leading to an overall performance decline.

\textbf{Feature Generator Evaluation.} The feature generator generates shared feature representations with cross-task discriminative ability for samples from two different tasks through adversarial training strategies. This section evaluates the impact of six commonly used feature generators in blockchain anomaly detection and tracing research on StealthLink's performance: ResNet-18, ResNet-50, ResNet-101 \cite{Resnet}, MLP, LSTM, and Transformer \cite{Transformer}. To ensure the reliability of the evaluation, we conducted a systematic test of the performance of all feature generators using ten-fold cross-validation, with results presented in Table \ref{tab:C5-encoder}.

\begin{table*}[htbp]
    \centering
    \caption{Performance Comparison of Different Feature Generators in StealthLink}
    \label{tab:C5-encoder}
    \resizebox{\textwidth}{!}{
        \begin{tabular}{cccccccc}
            \toprule
            \multirow{2}*{N} & \multirow{2}*{Evaluation Metric} & \multicolumn{6}{c}{Feature Generators} \\
            \cmidrule(lr){3-8}
            & & ResNet-18 & ResNet-50 & ResNet-101 & MLP & LSTM & Transformer \\
            \midrule

            \multirow{3}*{1}
            & Accuracy & 0.5772$\pm$0.1401 & 0.6672$\pm$0.2501 & 0.6419$\pm$0.0809 & 0.5996$\pm$0.0837 & 0.5296$\pm$0.0332 & 0.7879$\pm$0.1228 \\
            & Recall & 0.7638$\pm$0.2099 & 0.6759$\pm$0.1372 & 0.7426$\pm$0.1843 & 0.8163$\pm$0.2479 & 0.7127$\pm$0.2230 & 0.8324$\pm$0.0484 \\
            & F1-Score & 0.6575$\pm$0.1317 & 0.6715$\pm$0.1004 & 0.6886$\pm$0.0949 & 0.6914$\pm$0.0980 & 0.6076$\pm$0.0845 & \textbf{0.8096$\pm$0.0678} \\
            \midrule
            \multirow{3}*{3}
            & Accuracy & 0.6294$\pm$0.1333 & 0.8205$\pm$0.1732 & 0.6359$\pm$0.0772 & 0.7079$\pm$0.0820 & 0.5642$\pm$0.0556 & 0.9429$\pm$0.0199 \\
            & Recall & 0.7677$\pm$0.1387 & 0.6415$\pm$0.1846 & 0.8511$\pm$0.1129 & 0.7955$\pm$0.1009 & 0.8150$\pm$0.1809 & 0.9738$\pm$0.0167 \\
            & F1-Score & 0.6917$\pm$0.1206 & 0.7314$\pm$0.0279 & 0.7280$\pm$0.0301 & 0.7491$\pm$0.0392 & 0.6668$\pm$0.0485 & \textbf{0.9580$\pm$0.0118} \\
            \midrule
            \multirow{3}*{5}
            & Accuracy & 0.6552$\pm$0.1733 & 0.8915$\pm$0.1396 & 0.6933$\pm$0.0525 & 0.8192$\pm$0.0663 & 0.5857$\pm$0.0998 & 0.9657$\pm$0.0085 \\
            & Recall & 0.7584$\pm$0.1789 & 0.6830$\pm$0.1024 & 0.8658$\pm$0.0644 & 0.8528$\pm$0.0830 & 0.7367$\pm$0.2023 & 0.9660$\pm$0.0104 \\
            & F1-Score & 0.7030$\pm$0.1607 & 0.7735$\pm$0.0402 & 0.7688$\pm$0.0266 & 0.8357$\pm$0.0488 & 0.6526$\pm$0.0814 & \textbf{0.9657$\pm$0.0044} \\
            \midrule
            \multirow{3}*{10}
            & Accuracy & 0.7381$\pm$0.0381 & 0.9153$\pm$0.0874 & 0.6836$\pm$0.0684 & 0.8760$\pm$0.0433 & 0.6090$\pm$0.0830 & 0.9741$\pm$0.0031 \\
            & Recall & 0.8586$\pm$0.0581 & 0.6718$\pm$0.1024 & 0.8758$\pm$0.0608 & 0.9039$\pm$0.0755 & 0.7837$\pm$0.1513 & 0.9662$\pm$0.0036 \\
            & F1-Score & 0.7938$\pm$0.0279 & 0.7749$\pm$0.0615 & 0.7679$\pm$0.0550 & 0.8897$\pm$0.0355 & 0.6854$\pm$0.0282 & \textbf{0.9698$\pm$0.0015} \\
            \bottomrule
        \end{tabular}
    }
\end{table*}
\noindent \textbf{Transformer as a Feature Generator.} The Transformer demonstrates significant advantages as a feature generator when the sample size \(N \geq 3\). Its F1-score continues to improve as the data size increases, surpassing 90\% when \(N = 10\). This result validates the effectiveness of the Transformer in capturing shared features across tasks. This is because the Transformer can calculate global node association weights through multi-head attention mechanisms, enabling it to precisely capture implicit behavioral patterns in address interactions across different tasks. At the same time, the standard deviation of the Transformer decreases as the sample size grows, indicating that its performance stability improves significantly with the sufficiency of training data.

\noindent \textbf{Transfer Learning Paradigm Evaluation.} The transfer learning paradigm ensures the effective transfer of domain knowledge from the malicious account task to the cross-task migration of mixed coin transaction tracing. In this section, we compare models without the transfer learning paradigm, denoted as "w/o Transfer," with those employing different transfer learning paradigms, including: DAN (Domain Adversarial Networks) \cite{DAN}, DANN (Domain-Adversarial Neural Network) \cite{DANN}, and MCD (Maximum Classifier Discrepancy) \cite{MCD}. To ensure the reliability of the evaluation, we use ten-fold cross-validation to assess the performance of models with different feature generators. The results are shown in Table \ref{tab:C5-TLarchitecture}.

\begin{table*}[htbp]
	\centering
	\caption{Performance Comparison of Different Transfer Learning Frameworks in StealthLink}
	\label{tab:C5-TLarchitecture}
	\setlength{\tabcolsep}{6pt} 
	\begin{tabular}{cccccc}
		\toprule
		\multirow{2}*{N} & \multirow{2}*{Evaluation Metric} & \multicolumn{4}{c}{Transfer Learning Frameworks} \\
		\cmidrule(lr){3-6}
		& & DANN & DAN & MCD & w/o Transfer \\
		\midrule
		
		\multirow{3}*{1}
		& Accuracy & 0.7957$\pm$0.0691 & 0.6545$\pm$0.0993 & 0.7879$\pm$0.1288 & 0.5637$\pm$0.0675 \\
		& Recall & 0.6301$\pm$0.0726 & 0.6881$\pm$0.1590 & 0.8324$\pm$0.0484 & 0.7358$\pm$0.1470 \\
		& F1 Score & 0.7033$\pm$0.0115 & 0.6709$\pm$0.0536 & \textbf{0.8096$\pm$0.0678} & 0.6383$\pm$0.0523 \\
		\midrule
		\multirow{3}*{3}
		& Accuracy & 0.8483$\pm$0.0096 & 0.8510$\pm$0.0094 & 0.9429$\pm$0.0199 & 0.5275$\pm$0.0365 \\
		& Recall & 0.7347$\pm$0.0199 & 0.7617$\pm$0.0370 & 0.9738$\pm$0.0167 & 0.8106$\pm$0.1911 \\
		& F1 Score & 0.7879$\pm$0.0105 & 0.8041$\pm$0.0217 & \textbf{0.9580$\pm$0.0118} & 0.6391$\pm$0.0549 \\
		\midrule
		\multirow{3}*{5}
		& Accuracy & 0.8404$\pm$0.0116 & 0.8770$\pm$0.0088 & 0.9957$\pm$0.0085 & 0.5636$\pm$0.0423 \\
		& Recall & 0.7882$\pm$0.0118 & 0.8299$\pm$0.0223 & 0.9660$\pm$0.0104 & 0.7438$\pm$0.1711 \\
		& F1 Score & 0.8136$\pm$0.0080 & 0.8529$\pm$0.0103 & \textbf{0.9806$\pm$0.0044} & 0.6413$\pm$0.0339 \\
		\midrule
		\multirow{3}*{10}
		& Accuracy & 0.8675$\pm$0.0109 & 0.8775$\pm$0.0079 & 0.9741$\pm$0.0031 & 0.5512$\pm$0.0799 \\
		& Recall & 0.8989$\pm$0.0032 & 0.8984$\pm$0.0045 & 0.9662$\pm$0.0036 & 0.7733$\pm$0.1871 \\
		& F1 Score & 0.8829$\pm$0.0065 & 0.8878$\pm$0.0047 & \textbf{0.9698$\pm$0.0015} & 0.6436$\pm$0.0417 \\
		\bottomrule
	\end{tabular}
\end{table*}
As shown in Table~\ref{tab:C5-TLarchitecture}, the transfer learning framework significantly enhances the performance of StealthLink. Specifically, the "w/o Transfer" model performs significantly worse than the models using transfer learning frameworks in all sample settings. The F1 score is reduced by approximately $5\%$ to $30\%$ compared to the transfer learning models, proving that the transfer learning framework plays a crucial role in transferring domain knowledge from the malicious account task to the coin mixing transaction tracing task.

Among the transfer learning frameworks, MCD demonstrates the most significant effect on task knowledge transfer. Specifically, it achieves an F1 score of approximately 0.81 with only a small number of labeled samples (N=1), which is about $10\%$ higher than the F1 scores of the other two transfer learning frameworks. We speculate that this is because MCD, through the maximum classifier discrepancy strategy with dual classifiers, can effectively leverage domain knowledge from the malicious account detection task, generating clear and robust decision boundaries for classifying coin mixing accounts even when the cross-task feature space is not aligned.

\begin{figure}[htbp]
    \centering
    \begin{subfigure}[b]{0.45\textwidth}
        \includegraphics[width=\textwidth]{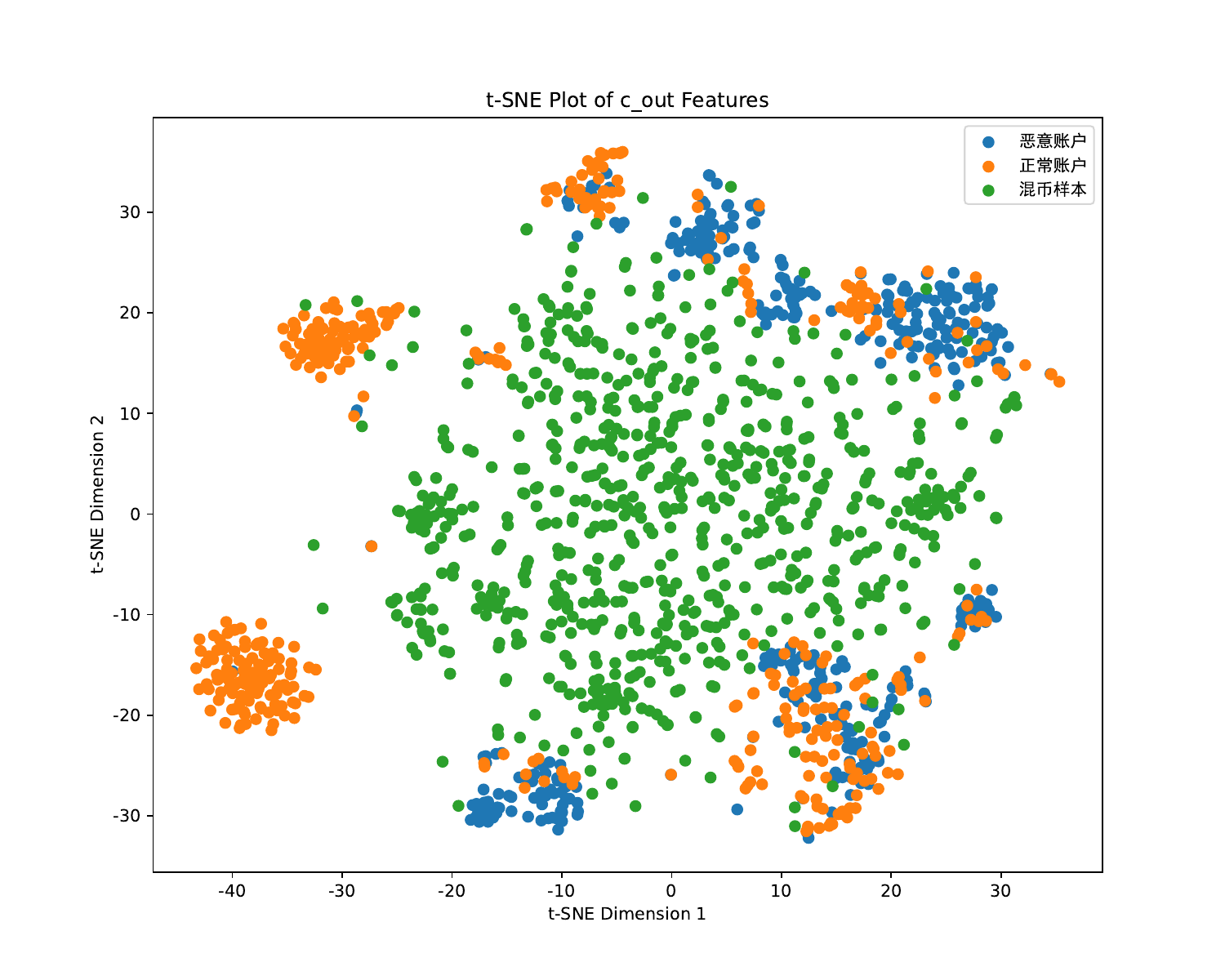}
        \caption{Visualization without Transfer Learning}
        \label{w/o Transfer visualization}
    \end{subfigure}
    \hfill
    \begin{subfigure}[b]{0.45\textwidth}
        \includegraphics[width=\textwidth]{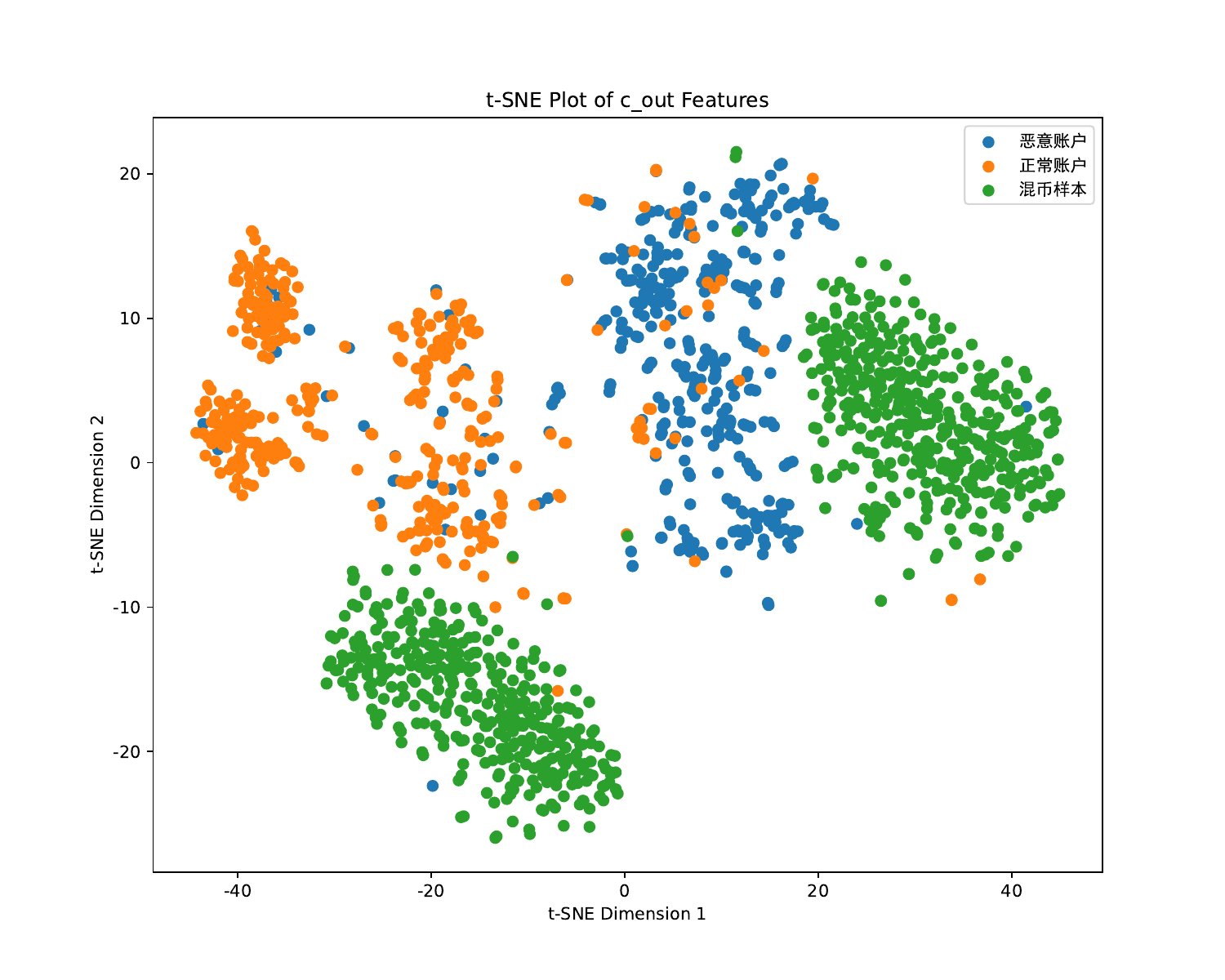}
        \caption{Visualization with MCD}
        \label{MCD visualization}
    \end{subfigure}
    \caption{Visualization of sample embeddings under the transfer learning framework.}
    \label{fig:C5-tSNE}
\end{figure}

To further evaluate the effectiveness of the transfer learning framework in cross-task knowledge transfer, this section uses the t-SNE dimensionality reduction algorithm to visually compare the high-dimensional latent feature space distributions of the MCD framework and the baseline model without transfer learning ("w/o Transfer"). The experimental data follows the cross-domain balance principle: 450 malicious accounts and 450 normal accounts are randomly selected from the BABD malicious account detection dataset to form the source domain samples, and an equal number of target domain samples are drawn from the Tornado Cash coin mixing transaction dataset. As shown in Figure~\ref{fig:C5-tSNE}, the blue, yellow, and green legends in the figure represent the latent feature distributions of malicious accounts, normal accounts, and coin mixing samples, respectively.

As illustrated in Fig.~\ref{fig:C5-tSNE}, the MCD-based transfer learning framework significantly enhances the separability of the coin mixing transaction tracing task, effectively achieving knowledge transfer from the malicious account detection task. Specifically, compared to the baseline model without transfer learning (w/o Transfer), the t-SNE visualization generated by the MCD framework reveals a pronounced spatial separation between clusters of malicious and benign accounts. Furthermore, coin mixing samples exhibit a linearly separable pattern along the main discriminative direction, indicating improved inter-class separability and intra-class compactness.
\begin{table*}[htbp]
    \centering
    \caption{Performance Comparison of Different Classifiers in StealthLink Framework}
    \label{tab:C5-Classifier}
    \resizebox{\textwidth}{!}{
    \begin{tabular}{ccccccccc}
        \toprule
        \multirow{2}*{N} & \multirow{2}*{Metric} & \multicolumn{7}{c}{Classifier} \\
        \cmidrule(lr){3-9}
        & & MLP & SVM & LR & KNN & RF & LSTM & XGBoost \\
        \midrule
        
        \multirow{3}*{1} 
        & Accuracy & 0.7879$\pm$0.1288 & 0.5884$\pm$0.0711 & 0.4990$\pm$0.0913 & 0.4990$\pm$0.0913 & 0.5393$\pm$0.0948 & 0.5850$\pm$0.1540 & 0 \\
        & Recall & 0.8324$\pm$0.0484 & 0.2937$\pm$0.1785 & 0.3585$\pm$0.2317 & 0.3584$\pm$0.2316 & 0.2950$\pm$0.1783 & 0.5371$\pm$0.2553 & 0 \\
        & F1 Score & \textbf{0.8096$\pm$0.0678} & 0.3918$\pm$0.0634 & 0.4172$\pm$0.0869 & 0.4172$\pm$0.0868 & 0.3814$\pm$0.0705 & 0.5600$\pm$0.1677 & 0 \\
        \midrule

        \multirow{3}*{3}
        & Accuracy & 0.9429$\pm$0.0199 & 0.4630$\pm$0.0568 & 0.4977$\pm$0.0548 & 0.4288$\pm$0.0556 & 0.5219$\pm$0.0551 & 0.6533$\pm$0.1095 & 0 \\
        & Recall & 0.9738$\pm$0.0167 & 0.3517$\pm$0.2707 & 0.6056$\pm$0.3216 & 0.4899$\pm$0.3504 & 0.6708$\pm$0.3023 & 0.5031$\pm$0.2837 & 0 \\
        & F1 Score & \textbf{0.9580$\pm$0.0118} & 0.3997$\pm$0.1640 & 0.5464$\pm$0.1734 & 0.4674$\pm$0.2172 & 0.5871$\pm$0.1680 & 0.5684$\pm$0.2485 & 0 \\
        \midrule

        \multirow{3}*{5}
        & Accuracy & 0.9957$\pm$0.0085 & 0.5028$\pm$0.0382 & 0.5461$\pm$0.0478 & 0.4807$\pm$0.0506 & 0.5618$\pm$0.0400 & 0.8647$\pm$0.0994 & 0.5279$\pm$0.0421 \\
        & Recall & 0.9660$\pm$0.0104 & 0.4010$\pm$0.2793 & 0.7407$\pm$0.2270 & 0.5398$\pm$0.3242 & 0.7959$\pm$0.1807 & 0.7330$\pm$0.1220 & 0.6298$\pm$0.2823 \\
        & F1 Score & \textbf{0.9806$\pm$0.0044} & 0.4462$\pm$0.1609 & 0.6288$\pm$0.1060 & 0.5085$\pm$0.1966 & 0.6588$\pm$0.0911 & 0.7934$\pm$0.1012 & 0.5279$\pm$0.0421 \\
        \midrule

        \multirow{3}*{10}
        & Accuracy & 0.9741$\pm$0.0031 & 0.5332$\pm$0.0266 & 0.5562$\pm$0.0439 & 0.5312$\pm$0.0346 & 0.5701$\pm$0.0310 & 0.9149$\pm$0.0731 & 0.5218$\pm$0.0356 \\
        & Recall & 0.9662$\pm$0.0036 & 0.4782$\pm$0.2714 & 0.7168$\pm$0.2575 & 0.6667$\pm$0.2936 & 0.8108$\pm$0.1649 & 0.8071$\pm$0.0848 & 0.7142$\pm$0.2868 \\
        & F1 Score & \textbf{0.9698$\pm$0.0015} & 0.5043$\pm$0.1567 & 0.6264$\pm$0.1323 & 0.5913$\pm$0.1529 & 0.6695$\pm$0.0791 & 0.8576$\pm$0.0782 & 0.5218$\pm$0.0356 \\
        \bottomrule
    \end{tabular}
    }
\end{table*}
\textbf{Classifier Evaluation.} During the model fine-tuning stage, the choice of classifier directly influences the association analysis of mixed transaction address pairs. This section evaluates the impact of seven widely-used classifiers—commonly adopted in blockchain anomaly detection and forensic studies—on the performance of the StealthLink framework. The classifiers include: Multi-Layer Perceptron (MLP), Support Vector Machine (SVM), Logistic Regression (LR), K-Nearest Neighbors (KNN), Random Forest (RF), Long Short-Term Memory network (LSTM), and eXtreme Gradient Boosting (XGBoost). To ensure evaluation reliability, we employ 10-fold cross-validation to systematically test the performance of all classifiers. Detailed experimental results are presented in Table~\ref{tab:C5-Classifier}.

From Table~\ref{tab:C5-Classifier}, it can be observed that the MLP classifier is more suitable for identifying the associations among coinjoin-related samples. Specifically, it achieves the highest F1-score under all sample settings. We hypothesize that this is because MLP leverages multiple layers of nonlinear activation functions to extract high-level feature combinations layer by layer, thereby capturing the complex fund flow patterns inherent in coin mixing transactions.

The experimental results indicate that linear classifiers (e.g., Support Vector Machine (SVM), Logistic Regression (LR)) and shallow models (e.g., K-Nearest Neighbors (KNN), Random Forest (RF)) exhibit relatively limited classification performance, with a maximum F1-score of only 67\%. We attribute this to the inherent limitations of linear decision boundaries in linear models, which make it difficult to effectively capture nonlinear patterns in coin mixing data—particularly in scenarios involving complex behaviors such as multiple deposits and withdrawals. Meanwhile, shallow classifiers lack the capacity for sufficient feature abstraction, which hampers their ability to perform global feature integration in high-dimensional transaction features.

It is noteworthy that the XGBoost method demonstrates a strong dependence on sufficiently labeled data. When $N=1$ or $N=3$, both the accuracy and recall of XGBoost are 0, which is due to the lack of training samples to determine effective splitting points. As the number of samples increases to 5 and 10, XGBoost begins to identify meaningful splits and shows a slight improvement in performance. This result highlights the limitations of tree-based models when dealing with extremely small training datasets.

\section{Conclusion}\label{sec:Conclusion}

In this chapter, we proposed \textbf{StealthLink}, a coin-mixing transaction tracing method based on cross-task invariant feature learning. The method introduces a coin-mixing subgraph fusion encoding module to construct an effective joint representation of mixed transactions, and employs a distribution discrepancy minimization strategy to enable knowledge transfer from malicious account detection to the domain of coin-mixing tracing. 

Extensive experiments on real-world datasets demonstrate that StealthLink achieves state-of-the-art discrimination performance while exhibiting strong few-shot learning capability and robustness to noisy pseudo-labeled data. 

In future work, we plan to extend our approach by incorporating cross-chain tracing mechanisms to explore behavioral patterns of malicious transactions that leverage coin mixing across different blockchains, thereby advancing the tracing of more covert malicious transaction activities.

\if CLASSOPTIONcaptionsoff
\newpage
\fi



\bibliographystyle{IEEEtran}
\bibliography{refs}

\end{document}